\documentclass[english,aps,prb,superscriptaddress,nofootinbib,letterpaper,preprint]{revtex4}
\usepackage[T1]{fontenc}
\usepackage[latin1]{inputenc}
\usepackage{geometry}
\usepackage{amsmath}
\usepackage{graphicx}
\usepackage{amssymb}

\usepackage{ae,aecompl}
\usepackage{subfigure}

\providecommand{\onlinecite}{\cite}

\usepackage{babel}
\makeatother
\begin{document}

\title{Tetratic Order in the Phase Behavior of a Hard-Rectangle System}

\author{Aleksandar Donev}

\affiliation{\emph{Program in Applied and Computational Mathematics}, \emph{Princeton
University}, Princeton NJ 08544}

\affiliation{\emph{PRISM}, \emph{Princeton University}, Princeton NJ 08544}

\author{Josh Burton}

\affiliation{\emph{Department of Physics}, \emph{Princeton University}, Princeton
NJ 08544}

\author{Frank H. Stillinger}

\affiliation{\emph{Department of Chemistry}, \emph{Princeton University}, Princeton
NJ 08544}

\author{Salvatore Torquato}

\email{torquato@electron.princeton.edu}

\affiliation{\emph{Program in Applied and Computational Mathematics}, \emph{Princeton
University}, Princeton NJ 08544}

\affiliation{\emph{PRISM}, \emph{Princeton University}, Princeton NJ 08544}

\affiliation{\emph{Department of Chemistry}, \emph{Princeton University}, Princeton
NJ 08544}

\begin{abstract}
Previous Monte Carlo investigations by Wojciechowski \emph{et al.}
have found two unusual phases in two-dimensional systems of anisotropic
hard particles: a tetratic phase of four-fold symmetry for hard squares
{[}\emph{Comp. Methods in Science and Tech., 10: 235-255, 2004}{]},
and a nonperiodic degenerate solid phase for hard-disk dimers {[}\emph{Phys.
Rev. Lett., 66: 3168-3171, 1991}{]}. In this work, we study a system
of hard rectangles of aspect ratio two, i.e., hard-square dimers (or
dominos), and demonstrate that it exhibits a solid phase with both
of these unusual properties. The solid shows tetratic, but not nematic,
order, and it is nonperiodic having the structure of a random tiling
of the square lattice with dominos. We obtain similar results with
both a classical Monte Carlo method using true rectangles and a novel
molecular dynamics algorithm employing rectangles with rounded corners.
It is remarkable that such simple convex two-dimensional shapes can
produce such rich phase behavior. Although we have not performed exact
free-energy calculations, we expect that the random domino tiling
is thermodynamically stabilized by its degeneracy entropy, well-known
to be $1.79k_{B}$ per particle from previous studies of the dimer
problem on the square lattice. Our observations are consistent with
a KTHNY two-stage phase transition scenario with two continuous phase
transitions, the first from isotropic to tetratic liquid, and the
second from tetratic liquid to solid.
\end{abstract}
\maketitle

\newcommand{\Cross}[1]{\left|\mathbf{#1}\right|_{\times}}

\newcommand{\CrossL}[1]{\left|\mathbf{#1}\right|_{\times}^{L}}

\newcommand{\CrossR}[1]{\left|\mathbf{#1}\right|_{\times}^{R}}

\newcommand{\CrossS}[1]{\left|\mathbf{#1}\right|_{\boxtimes}}

\newcommand{\V}[1]{\mathbf{#1}}

\newcommand{\M}[1]{\mathbf{#1}}

\newcommand{\D}[1]{\Delta#1}

\newcommand{\sV}[1]{\boldsymbol{#1}}

\newcommand{\sM}[1]{\boldsymbol{#1}}

\newcommand{\grad}{\boldsymbol{\nabla}}

\newcommand{\periodcite}[1]{.\cite{#1} }
\newcommand{\commacite}[1]{,\cite{#1} }
\newcommand{\spacecite}[1]{\cite{#1} }

\section{Introduction}

Hard-particle systems have provided a simple and rich model for investigating
phase behavior and transport in atomic and molecular materials. It
is long-known that a pure hard-core exclusion potential can lead to
a variety of behaviors depending on the degree of anisotropy of the
particles, including the occurrence of isotropic and nematic liquids,
layered smectic, and ordered solid phases\periodcite{SimpleComplex_Liquids}
Through computer investigations of various particle shapes, other
phases have been found, such as the biaxial\spacecite{Biaxial_Ellipsoid_Fluid}
(recently synthesized in the laboratory\spacecite{BiaxialNematic_Boomerang})
and cubatic phases in three dimensions, in which the axes of symmetry
of the individual particles align along two or three perpendicular
axes (directors). One only need look at simple shapes in two dimensions
to discover interesting phases. In recent work, Wojciechowski \emph{et
al.} studied hard squares and found the first example of a tetratic
liquid phase at intermediate densities\periodcite{HardSquares_MC} In
a tetratic liquid, there is (quasi) long-range orientational ordering
along two perpendicular axes, but only short-range translational ordering.
The solid phase is the expected square lattice, with quasi-long-range
periodic ordering. On the other hand, by studying hard-disk dimers
(two disks fused at a point on their boundary), they have identified
the first example of a nonperiodic solid phase at high densities\periodcite{DiskDimers_Solid}
In this phase, the centroids of the particles are ordered on the sites
of a triangular lattice. However, the orientations of the dimers are
disordered, leading to a high degeneracy entropy of the nonperiodic
solid and a lower free energy as compared to periodic solids.

In this paper, we look at systems of rectangles of aspect ratio $\alpha=a/b=2$,
i.e., hard-square dimers (or dominos). Since the aspect ratio is far
from unity, it is not clear \emph{a priori} whether nematic or tetratic
orientational ordering (or both) will appear. Recent Density Functional
Theory calculations\commacite{HardRodFluids_DFT} extending previous
work based on scaled-particle theory\commacite{HardRodFluids_SPT} have
predicted that for $\alpha=2$ the tetratic phase is only metastable
with respect to the ordered solid phase in which all particles are
aligned. However, these calculations are only approximate and the
authors point out that tetratic order is still possible in spatially
ordered phases. An obvious candidate for forming a stable tetratic
phase are dominos: two dominos paired along their long edges form
a square, and these squares can then form a square lattice assuming
one of two random orientations, thus forming a tetratic phase with
degeneracy entropy of $\ln(\sqrt{2})$. In fact, one does not need
to pair up the rectangles but rather simply tile a square lattice
with dominos which randomly assume one of the two preferred perpendicular
directions. The degeneracy entropy of this domino tiling has been
calculated exactly to be $(2G/\pi)k_{B}\approx0.58313k_{B}$\commacite{DimerCovering_square_1,DimerCovering_square_2}
where $G=\sum_{n=0}^{\infty}{(-1)^{n}(2n+1)^{-2}}\approx0.91597$
is Catalan's constant. At high densities, free-volume theory\spacecite{SimpleComplex_Liquids}
predicts that the configurational entropy (per particle) diverges
like \[
S_{\textrm{FV}}\sim f\ln(1-\phi/\phi_{c})+S_{\textrm{conf}},\]
where $f$ is the (effective) number of degrees of freedom per particle,
$\phi_{c}$ is the volume fraction (density) at close packing, and
$S_{\textrm{conf}}$ is an additive constant due to collective exclusion-volume
effects. Therefore, the densest solid is thermodynamically favored,
but if several solids have the same density the additive factor matters.
Therefore, for hard rectangles, for which the maximal density is $\phi_{c}=1$
and is achieved by a variety of packings, the degeneracy entropy can
dominate $S_{\textrm{conf}}$ and thus the nonperiodic random tiling
can be thermodynamically favored. Indeed, we demonstrate that our
simulations of the hard-domino system produce high-density phases
with structures very similar to that of a random covering of the square
lattice with dimers.

The phase transitions in two-dimensional systems are of interest to
the search for continuous KTHNY\spacecite{KT_Transition_2D,HN_Transition_2D,Young_Transition_2D}
transitions between the disordered liquid and the ordered solid phase.
At present there is no agreement on the nature of the transition even
for the hard-disk system. A previous study of the melting of a square-lattice
crystal, stabilized by the addition of three-body interactions, found
evidence of a (direct) first-order melting\periodcite{SquareCrystal_3body}
Our observations for the domino system are relatively consistent with
a KTHNY-like two-stage transition: a continuous phase transition from
an isotropic to a tetratic liquid with (quasi-) long-range tetratic
order around $\phi\approx0.7$, and then another continuous transition
from tetratic liquid to tetratic solid with quasi-long-range translational
order $\phi\approx0.8$. However, we cannot rule out the possibility
of a weak first-order phase transition between the two phases without
more detailed simulations.

This paper is organized as follows. In Section \ref{Section_Simulations},
we present the simulation techniques used to generate equilibrated
systems at various densities. In Section \ref{Section_Results}, we
analyze the properties of the various states, focusing on the orientational
and translational ordering in the high-density phases. We conclude
with a summary of the results and suggestions for future work in Section
\ref{Section_Conclusions}.

\section{\label{Section_Simulations}Simulation Techniques}

In this section, we provide additional details on the MC and MD algorithms
we implemented. It is important to point out that it is essential
to implement techniques for speeding up the near-neighbour search,
in both MC and MD. For rectangles with a small aspect ratio, we employ
the well-known technique of splitting the domain of simulation into
cells (bins) larger than a particle diameter $D=\sqrt{a^{2}+b^{2}}$,
and consider as neighbors only particles whose centroids belong to
neighboring cells. Additional special techniques more suitable for
very aspherical particles or systems near jamming are described in
Ref. \onlinecite{Event_Driven_HE}.

\subsection{\label{Section_MC}Monte Carlo}

We have implemented a standard MC algorithm in the $NVT$ ensemble,
with the additional provision of changing the density by growing or
shrinking the particles in small increments. Each rectangle is described
by the location of its centroid ($x,y$) and orientation $\theta$.
For increased computational speed the pair $(\sin\theta,\cos\theta)$
may be used to represent the orientation. In a trial MC step, a rectangle
is chosen at random and its coordinates are changed slightly, either
translationally ($\Delta x,\Delta y$) or orientationally ($\Delta\theta$).
Every move has an equal chance of being translational or orientational.
The rectangle's new position is then compared against nearby rectangles
for overlap; if there is no overlap, the trial move is accepted. We
call a sequence of $N$ trials a cycle. The simulation evolves through
\textit{stages}, defined by a speed $n_{cycles/stage}$. At the end
of a cycle, pressure data are collected by the virtual-scaling method
of Eppenga and Frenkel\periodcite{VirtualScaling_MC} Namely, $p=PV/NkT=1+\phi\alpha/2$
where $\alpha$ is the rate at which growing the particles causes
overlaps. At the end of a stage, order parameters and other statistics
are collected, and then the packing fraction $\phi$ is changed by
a small value $\D{\phi}$; it may be increased, decreased, or not
changed at all. If $\D{\phi}>0$, then $\phi$ cannot necessarily
change by $\D{\phi}$ every stage, because the increase could create
overlaps. We scale down the increase by factors of 2 until a $\D{\phi}_{\textrm{eff}}$
is found that does not cause any overlaps when applied. Typical values
for runs are $n_{cycles/stage}=1000$ and $\D{\phi}=\pm1\times10^{-5}$.
Since there is a limit on how fast one can increase the density in
such a Monte Carlo simulation, especially at very high densities,
we use Molecular Dynamics to compress systems to close packing.

\begin{figure}
\begin{center}\includegraphics[%
  width=0.45\linewidth,
  keepaspectratio]{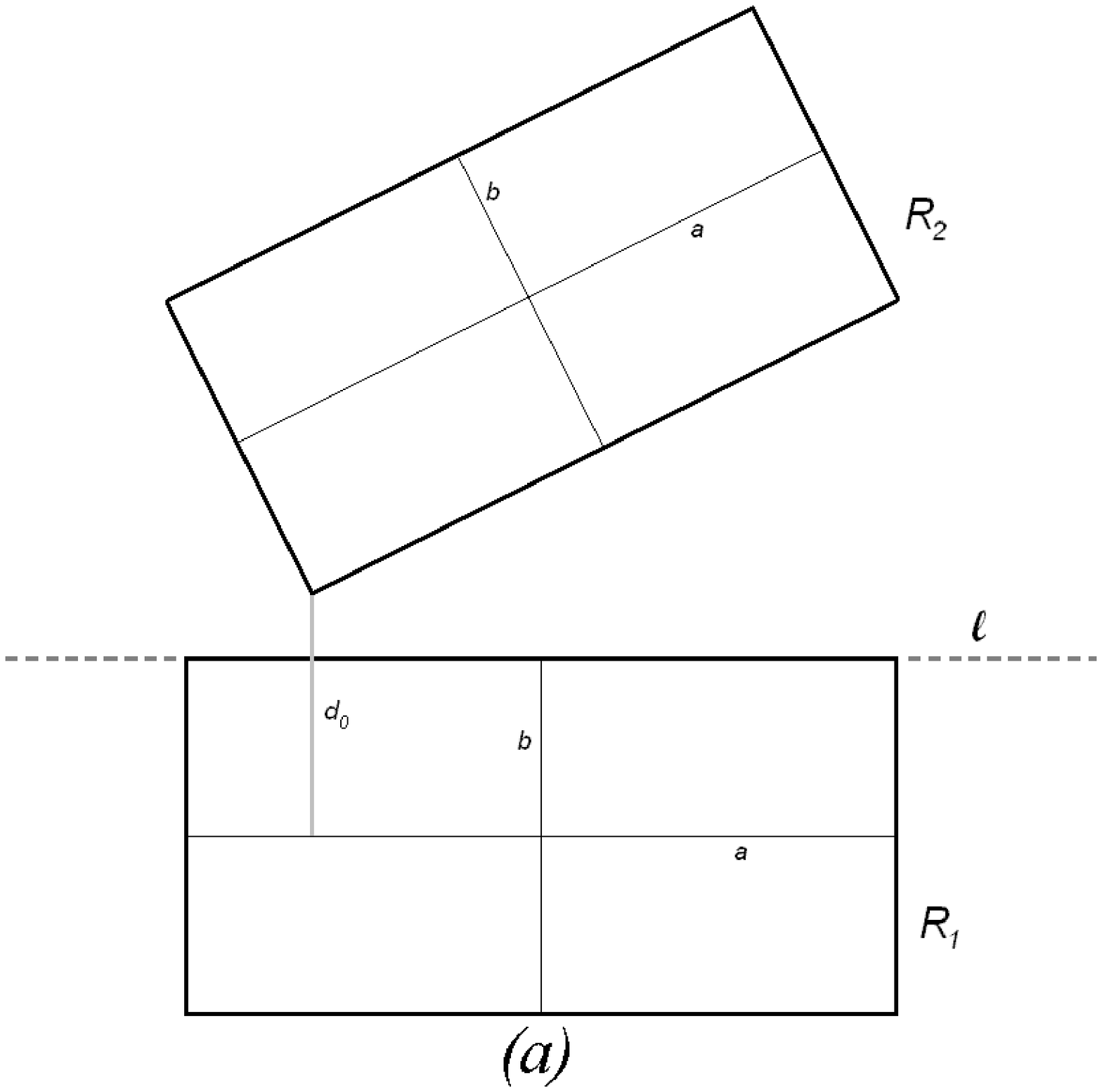}\includegraphics[%
  width=0.45\linewidth,
  keepaspectratio]{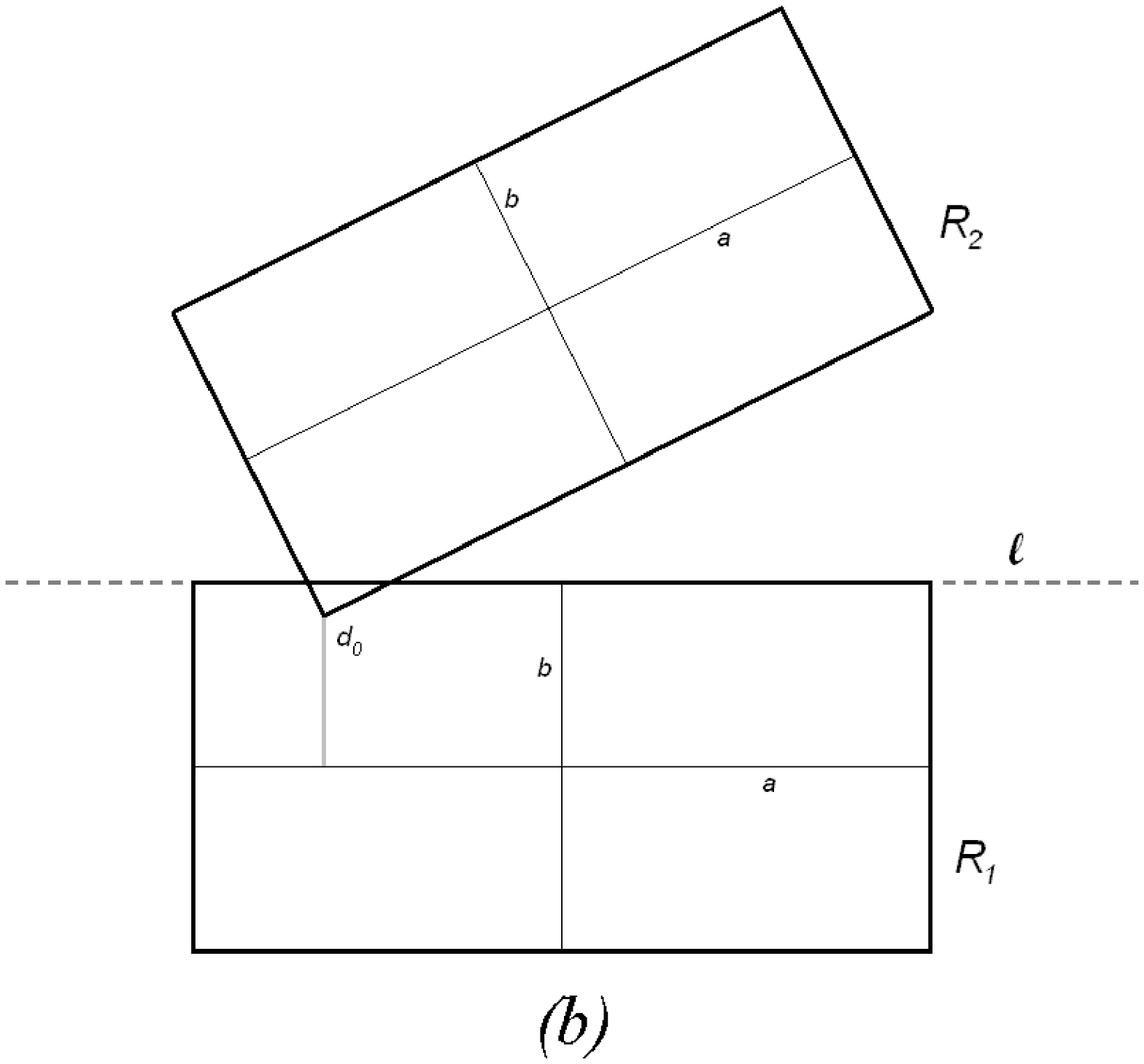}\end{center}

\caption{\label{RectangleOverlap}Illustration of the optimized overlap test
for two rectangles. The axes are $\bar{a}$ and $\bar{b}$, with semiaxes
$a$ and $b$, and the length from $\bar{a}$ to the closest corner
of $R_{2}$ is $d_{0}$. (\emph{Left)} If $d_{0}\geq b$, then the
rectangles do not overlap. (Right) If $d_{0}<b$, then the rectangles
overlap.}
\end{figure}

The overlap test is by far the largest computational bottleneck in
the MC program. The overlap test for two rectangles is based on the
following fact: Two rectangles $R_{1}$ and $R_{2}$ do not overlap
if and only if a separating line $\ell$ can be drawn such that all
four corners of $R_{1}$ lie on one side of the line and all four
corners of $R_{2}$ lie on the other side\periodcite{OBB_Thesis} The
corners of both rectangles are allowed to coincide with $\ell$. Without
loss of generality, we may assume $\ell$ is drawn parallel to one
of the rectangles' major axes and runs exactly along that rectangle's
side. The problem of testing all possible lines $\ell$ thus reduces
to testing the eight lines that coincide with the edges of $R_{1}$
and $R_{2}$. The test can be optimized somewhat further, as illustrated
in Fig. \ref{RectangleOverlap}. An axis $\bar{a}$ of $R_{1}$ is
chosen. The distance from the center of $R_{2}$ to $\bar{a}$ is
found. Then the distance $d_{0}$ of closest approach of $R_{2}$
to $a$ is found by subtracting a sine and a cosine; this distance
corresponds to the corner of $R_{2}$ that is closest to $\bar{a}$.
By comparing $d_{0}$ with the length $b$ of the other semiaxis of
$R_{1}$, two possible lines $\ell$, corresponding to two opposite
sides of $R_{1}$, can be tested at once. If $d_{0}<b$, there is
an overlap. In this way, four different values of $d_{0}$ are calculated;
one for each axis of each rectangle. If no comparison finds an overlap,
there is no overlap.

\subsection{\label{Section_MD}Molecular Dynamics}

\begin{figure}
\begin{center}\includegraphics[%
  width=0.5\columnwidth,
  keepaspectratio]{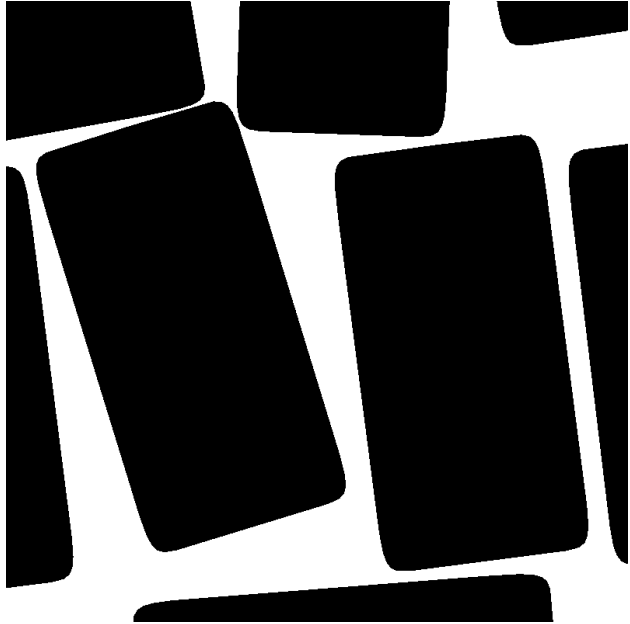}\end{center}

\caption{\label{SHE.alpha=3D2.0.exp=3D7.5.example}An snapshot of a few superellipses
(exponent $\zeta=7.5$) used in the MD simulations. It can be seen
that the particle shape is very close to a rectangle.}
\end{figure}

MC simulations are typically the most efficient when one is only interested
in stable equilibrium properties. We have previously developed a molecular
dynamics (MD) algorithm aimed at studying hard nonspherical particles
and applied it to systems of hard ellipses and ellipsoids\periodcite{Event_Driven_HE}
We have since generalized the implementation to also handle {}``superellipses''
and {}``superellipsoids'', which are generalized smooth convex shapes
capable of approximating centrally symmetric shapes with sharp corners
such as rectangles. A superellipse with semiaxes $a$ and $b$ is
given by the equation\[
\left[\left|\frac{x}{a}\right|^{2\zeta}+\left|\frac{y}{b}\right|^{2\zeta}\right]^{1/\zeta}\leq1,\]
where $\zeta\geq1$ is an exponent. We add an exponent $1/\zeta$
above in order to properly normalize the convex function defining
the particle shape, even though it is not strictly necessary. When
$\zeta=1$ we get the simple ellipse, and when $\zeta\rightarrow\infty$
we obtain a rectangle with sides $2a$ and $2b$. The higher the exponent
the sharper the corners become. The smoothing of the corners of the
rectangle enables us to apply our collision-driven MD algorithm\commacite{Event_Driven_HE}
with few changes from the case of ellipses. Details of this implementation
will be given elsewhere. The floating-point cost of the algorithm
increases as the exponent increases, while the numerical stability
decreases. We have used an exponent $\zeta=7.5$ for the studies presented
here (for this exponent the ratio of the areas of the superellipse
and the true rectangle is $0.9934$). Figure \ref{SHE.alpha=3D2.0.exp=3D7.5.example}
gives an illustration of the particle shape.

There are some advantages of the MD simulation over MC. The shapes
of the particles can change arbitrarily fast in an easily controlled
manner by simply adding a dynamic growth rate $\gamma=da/dt=\alpha db/dt$.
If $\gamma>0$, i.e., the density is increasing, two colliding particles
simply get an extra repulsive boost that ensures no overlaps are created.
The velocities are periodically rescaled to $T=1$ to compensate for
the induced heating or cooling due to the particle growth\periodcite{Event_Driven_HE}
In general, (common) MC methods do not work well near close packing,
while MD methods, especially event-driven ones, can successfully be
used to study the neighbourhood of jamming points. Additionally, pressure
measurement is more natural in the MD method, as the pressure can
be directly obtained from time averages of the momentum exchange in
binary collisions between particles. We have found this pressure measurement
to be much more precise than using virtual particle scaling in MC
simulations.

\section{\label{Section_Results}Results}

By using either the MC or the MD algorithm with small particle growth
rate ($\D{\phi}$ or $\gamma$), we have traced the (quasi)equilibrium
phase behavior of systems of dominos over a range of densities. In
this section, we present several techniques for measuring orientational
and translational order for a given configuration of particles, as
well as the results of such measurements for the generated states.
We have tested our codes by first applying them to hard squares and
comparing the results to those in Ref. \onlinecite{HardSquares_MC}, and
we have observed good quantitative agreement throughout. Our MC pressure
measurement systematically slightly underestimates the pressure compared
to the $NPT$ ensemble used in Ref. \onlinecite{HardSquares_MC} and to
our MD simulations. We present some of the results for the MC, and
others for the MD simulations, marking any quantitative differences.
The two techniques always produced qualitatively identical results. 

Describing the statistical properties of the observed states would
require specifying all of the $n$-particle correlation functions.
The most important is the pair correlation function $g_{2}(r,\psi,\Delta\theta)$.
Given a particle, $g_{2}(r,\psi,\Delta\theta)$ is the probability
density of finding another particle whose centroid is a distance $r$
away (from the centroid of the particle), at a displacement angle
of $\psi$ (relative to the first particle's coordinate axes), and
with a orientation of $\Delta\theta$ (relative to the particle's
orientation). The normalization of $g_{2}$ is such that it is identically
unity for an ideal gas. We will use an equivalent representation where
we fix a particle at the origin such that the longer rectangle axis
is along the $x$ axes, and represent pair correlations with $g_{2}(\D{x},\D{y},\D{\theta})$,
giving the probability density that there is another particle whose
centroid is at position $(\D{x},\D{y})$ and whose major axis is at
a relative angle of $\D{\theta}$. Since a three-dimensional function
is rather difficult to calculate accurately and visualize, we can
separate the translational and orientational components and average
over some of the dimensions to reduce it to a one- or two-dimensional
function.

\subsection{\label{Section_EOS}Equation Of State}

The pressure as a function of density can be most accurately measured
in the MD simulations. There is no exact theory that can predict the
entire equation of state (EOS) for a given many-particle system. However,
there are two simple theories that produce remarkably good predictions
for a variety of systems studied in the literature. For the isotropic
fluid (gas) phase of a system of hard dominos, scaled-particle theory
(SPT)\spacecite{SPT_convex_2D} predicts \begin{equation}
p=\frac{1}{1-\phi}+\frac{9}{2\pi}\frac{\phi}{(1-\phi)^{2}},\label{EOS_SPT}\end{equation}
and modifications to account for possible orientational ordering are
discussed in Refs. \onlinecite{HardRodFluids_SPT,HardRodFluids_DFT}. For
the solid phase, the free-volume (FV) theory predicts a divergence
of the pressure near close packing of the form\begin{equation}
p=\frac{3}{1-\phi/\phi_{c}},\label{EOS_FV}\end{equation}
and (liquid-state) density functional theory can be used to make quantitative
predictions at intermediate densities\periodcite{HardRodFluids_DFT} For
superellipses with exponent $\zeta=7.5$ the maximal density is somewhat
less than $1$ and we take it to be equal to the ratio of the areas
of the particle and a true rectangle, $\phi_{c}\approx0.9934$.

\begin{figure}
\begin{center}\includegraphics[%
  width=0.95\columnwidth,
  keepaspectratio]{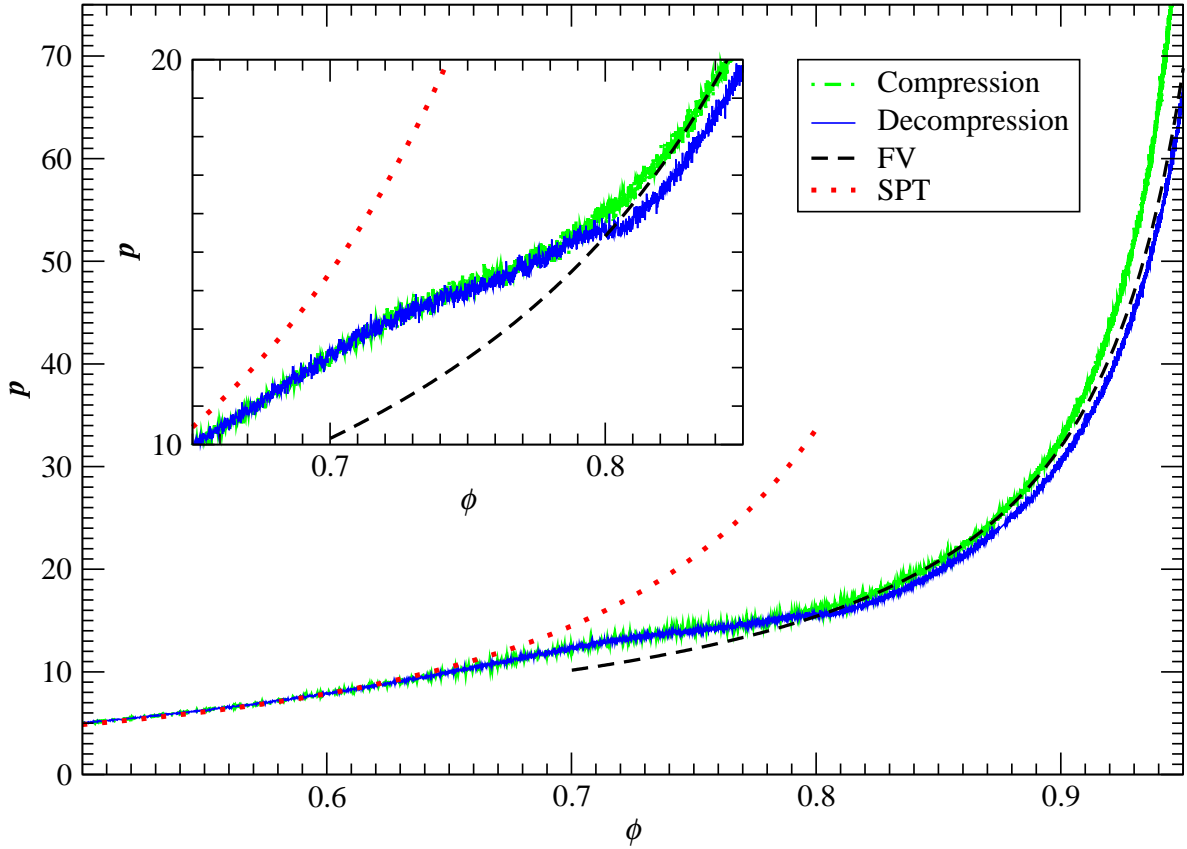}\end{center}

\caption{\label{PV_dominos}(Color online) Reduced pressure $p=PV/NkT$ in
a system of $N=5000$ superellipses with exponent $\zeta=7.5$ during
MD runs with $\gamma=\pm2.5\cdot10^{-5}$. The predictions of simple
versions of SPT and FV theory are also shown for comparison. The agreement
with FV predictions is not perfect; a numerical fit produces a coefficient
$2.9$ instead of $3$ in the numerator of Eq. (\ref{EOS_FV}). Particularly
noticeable are the change in slope around $\phi\approx0.72$ and also
the transition onto a solid branch well-described by free-volume approximation
around $\phi\approx0.8$. Starting the decompression from an ordered
tiling in which all rectangles are aligned produces identical pressure
to within the accuracy available. Systems of $N=1250$ and $N=10000$
particles, as well as a wide range of particle growth rates, were
investigated to ensure that there were no strong finite-size or hysteresis
effects. In faster compressions of an isotropic liquid one gets smaller
final densities due to the occurrence of defects such as vacancies
or grain boundaries.}
\end{figure}

The numerical EOS from the $NVT$ MD simulation are shown in Fig.
\ref{PV_dominos} for both a slow compression starting from an isotropic
liquid and a decompression starting from a perfect random domino tiling
generated with the help of random spanning trees, as explained in
Ref. \onlinecite{RandomTiling_SpanningTree}. We note that the random domino
tiling used was generated inside a square box (see Fig. \ref{CompareDomino_SHE})
even though periodic boundary conditions were used in the actual simulation.
We expect this to have a very small effect\periodcite{RandomTilings_Entropy}
It is clearly seen from the figure that there is a transition from
the liquid to the solid branch in the region $\phi\approx0.7$ and
$\phi\approx0.8$, although no clear discontinuities or a hysteresis
loop are seen (which would be indicative of a first-order phase transition).
Compressing an isotropic liquid invariably freezes some defects and
thus the jamming density is smaller (and the pressure is thus higher)
than in the perfect crystal.

\subsection{\label{Section_S_4}Orientational Order}

Orientational order can be measured via the orientational correlation
function of order $m$\begin{equation}
G_{m}(r)=\langle\cos(m\Delta\theta)\rangle_{r},\end{equation}
 where $m$ is an integer and the average is taken over all pairs
of particles that are at a distance between $r$ and $r+dr$ apart
from each other. The one-dimensional function $G_{m}(r)$ can be thought
of as giving normalized Fourier components of the distribution of
relative orientations versus interparticle distance. When $m=2$,
it measures the degree of nematic ordering (parallel alignment of
the particles' major axes), and when $m=4$ it measures the degree
of tetratic ordering (parallel alignment of the particles' axes).
The infinite-distance value $\underset{r\rightarrow\infty}{\lim}G_{m}(r)=S_{m}$
gives a scalar measure of the tendency of the particles to align with
a global coordinate system; $S_{2}$ is the usual nematic order parameter,
and $S_{4}$ is the tetratic order parameter. They can be very easily
calculated from an alternative definition

\begin{equation}
S_{m}=\underset{\theta_{0}}{\max}\left\langle \cos[m(\theta-\theta_{0})]\right\rangle ,\label{S_m_definition}\end{equation}
which can be converted into an eigenvalue problem (in any dimension)
for the case $m=2$\periodcite{BiaxialOrderMetrics} When $m=4$, we can
rewrite it in the same form as $m=2$ by replacing $\theta$ with
$2\theta$. The vector $\V{n}_{m}=(\cos\theta_{0},\sin\theta_{0})$
determines a natural coordinate system for orientationally ordered
phases. It is commonly called the director for nematic phases ($m=2$),
and we will refer to it as a bidirector for tetratic phases ($m=4$).

In two-dimensional liquid-crystalline phases, it is expected that
there can be no long-range orientational ordering, but rather only
quasi-long-range orientational ordering\periodcite{LRO_HardRods_2D} Based
on elasticity theory with a single renormalized Frank's constant $\tilde{K}=\pi K/(8k_{B}T$),
it is predicted\spacecite{SpherocyllindersPhases_2D} that there will
be a power-law decay of the correlations at large distances, $G_{m}(r)\sim r^{-\eta}$,
where\begin{equation}
\eta=m^{2}/16K.\label{eta_K}\end{equation}
This would imply that $S_{m}$ vanishes with increasing system size,\begin{equation}
S_{m}\sim N^{-\eta/4}.\label{S_m_decay}\end{equation}
We note that this prediction is based on literature for the nematic
phase. We are not aware of any theoretical work explicitly for a tetratic
phase.

The KTHNY theories predict that the isotropic liquid first undergoes
a defect-mediated second-order transition into an orientationally
quasi-ordered but translationally disordered state when $\tilde{K}=1$
by disclination pair binding. At higher densities there is another
second-order phase transition into a solid that has long-range orientational
order and quasi-long range translational order, mediated by dislocation
pair binding. The validity of this theory is still contested even
for hard disks\commacite{HardDiskMelting_Scaling} and its applicability
to systems where there is strong coupling between orientational and
translational molecular degrees of freedom is questionable. Additionally,
the basic theory needs to be modified to include three independent
elastic moduli as opposed to only two in the case of six-fold rotational
symmetry.

\begin{figure}
\begin{center}\includegraphics[%
  width=0.95\columnwidth,
  keepaspectratio]{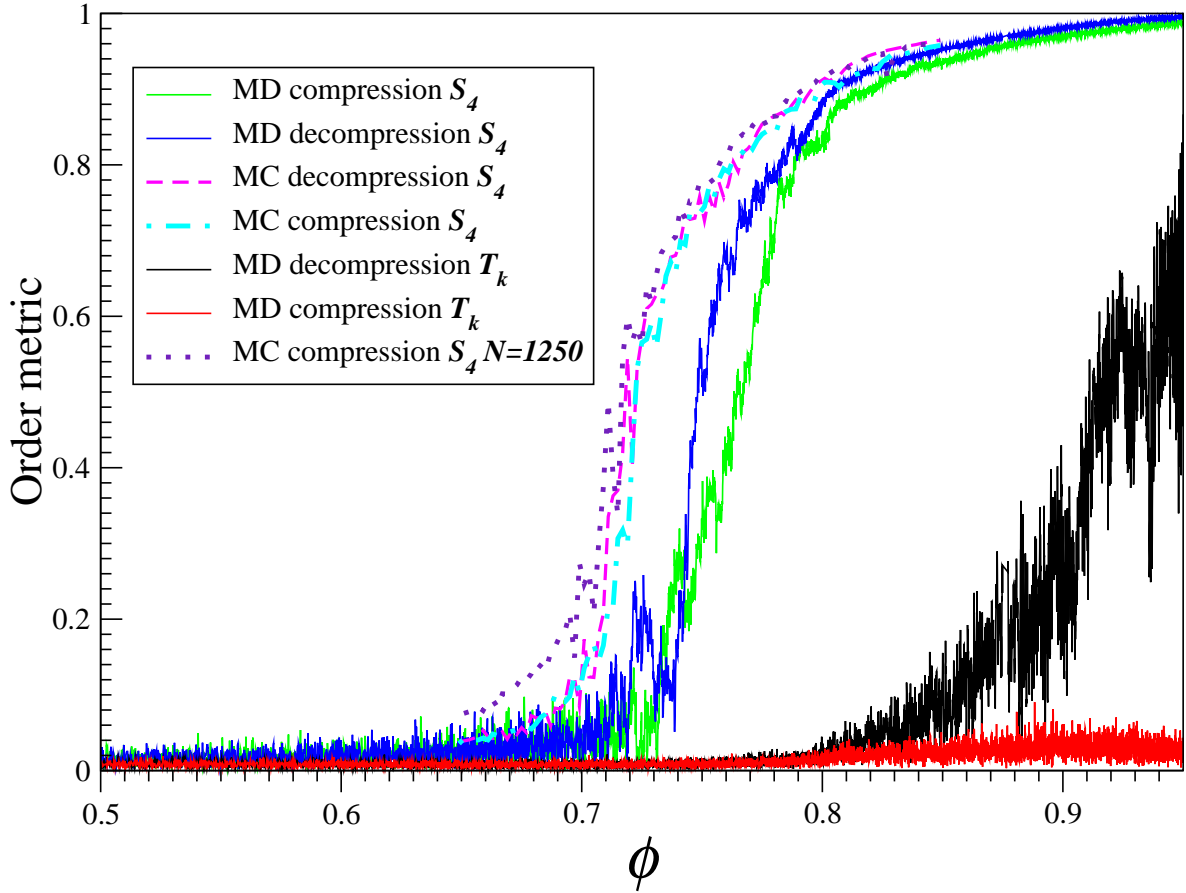}\end{center}

\caption{\label{S4.domino.hysteresis}(Color online) Values of the order metrics
$S_{4}$ {[}c.f. Eq. (\ref{S_m_definition}){]} and $T_{k}$ {[}c.f.
Eq. (\ref{S_k_scalar}){]} for snapshot configurations along (de)compression
MC and MD runs with $N=5000$ particles. Also shown is an MC run with
$N=1250$ for comparison. A transition in $S_{4}$ is visible around
$\phi\approx0.7$, and around $\phi\approx0.8$ for $T_{k}$. The
values of $T_{k}$ are much smaller and variable for compression runs
due to the sensitivity to the exact axes used, however, zooming in
reveals a qualitative change in $T_{k}$ around $\phi\approx0.8$
even in the compression data.}
\end{figure}

The observed change in $S_{4}$ as an isotropic liquid is slowly compressed
is shown in Fig. \ref{S4.domino.hysteresis} for both MD and MC runs.
It is clearly seen that tetratic order appears in the system around
$\phi\approx0.7$ and increases sharply as the density is increased,
approaching perfect order ($S_{4}=1$) at close packing. Throughout
this run $S_{2}$ remains close to zero and thus no spontaneous nematic
ordering is observed. It is important to note that superellipsoids
are not perfect rectangles and have rounded sides. It is therefore
not unexpected that they show less of a tendency toward tetratic (right-angle)
ordering, and have the isotropic-tetratic (IT) transition at slightly
higher densities. Additionally, the MD runs show more (correlated)
variability due to the strong correlations between successive states
(snapshots). Therefore, we prefer to consider the MC results, other
than at very high densities when we have to resort to MD studies.
We have also performed runs decreasing the density of a random domino
tiling, which has no nematic but has perfect tetratic order, and the
resulting $S_{4}$ is also shown in the figure. Only a mild hysteresis
is seen, especially for the MC runs, which would be indicative of
a continuous IT transition, or at least a weakly discontinuous one.

\begin{figure}
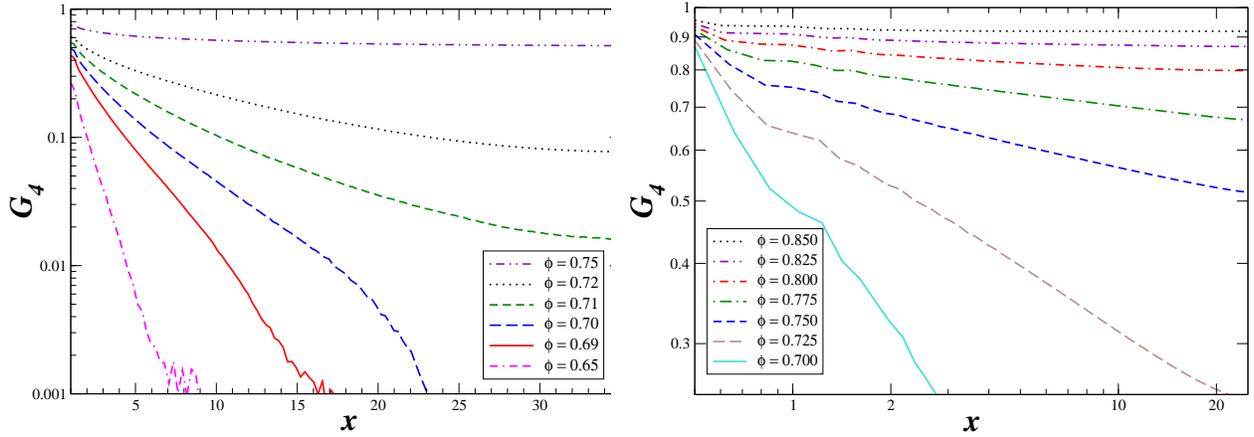

\begin{center}\includegraphics[%
  width=0.5\textwidth,
  keepaspectratio]{5_home_HPC_Packing_Papers_Dominos_graphics_G4_domino_N=10000_MC.eps}\includegraphics[%
  width=0.5\textwidth,
  keepaspectratio]{6_home_HPC_Packing_Papers_Dominos_graphics_G4_domino_N=5000_MC.eps}\end{center}

\caption{\label{G4.domino.MC}(Color online) \emph{Left:} Log-linear plot
of $G_{4}(x)$ for thoroughly equilibriated samples of $N=10000$
particles, showing the decay of orientational ordering with distance
$x=r/D$. The isotropic-tetratic transition occurs between $\phi=0.69$
and 0.70, when the tail behavior of $G_{4}(r)$ changes from exponential
(short-ranged) to slower-than-exponential. \emph{Right:} Log-log plot
of $G_{4}(x)$ for equilibrated systems of $N=5000$ particles, showing
power-law decay indicative of quasi-long-range tetratic order. The
fitted values of the power-law exponent are shown in Fig. \ref{G4.domino.MC.power}.}
\end{figure}

\begin{figure}
\begin{center}\includegraphics[%
  width=0.6\textwidth,
  keepaspectratio]{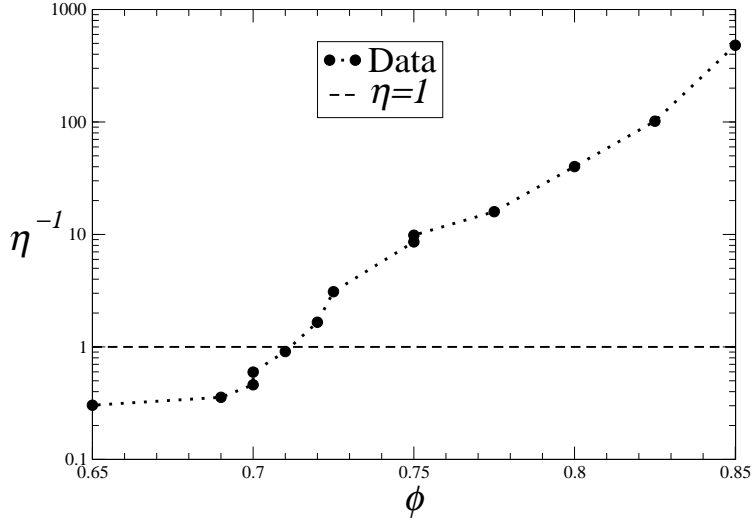}\end{center}

\caption{\label{G4.domino.MC.power}Log-linear plot of $1/\eta$, where $\eta$
is the exponent of decay of $G_{4}(r)$ found by fitting the $G_{4}(x)$
data in Fig. \ref{G4.domino.MC} to a power-law curve, $G_{4}(x)=Cx^{-\eta}$.}
\end{figure}

Figure \ref{G4.domino.MC} shows $G_{4}(r)$ for a collection of states
in the vicinity of the IT transition, thoroughly equilibriated using
MC, on both a log-linear (lower densities) and a log-log (higher densities)
scale. It is seen that there is a clear change in the long-range behavior
of $G_{4}(r)$ as the density crosses above $\phi_{c}\approx0.70$,
from an exponential decay typical of an isotropic liquid, to a slower-than
exponential decay at higher densities. The decay tails at higher densities
are rather consistent with a power-law decay, and the fitted exponents
$\eta$ are shown in Fig. \ref{G4.domino.MC.power}. It can be seen
that $\eta$ crosses the value $\eta_{c}=1$ predicted by KTHNY theory
when $\phi\approx0.71$, which is very consistent with the estimates
of the location of the IT transition through the other methods above.
It is not clear to us why the authors of Ref. \onlinecite{HardSquares_MC}
used the value of the exponent predicted by KTHNY theory for the bond-bond
orientational order in the hard-disk system, $\eta_{c}=1/4$, instead
of $\eta_{c}=1$. The somewhat higher values for $S_{4}$ for the
system with $N=1250$ relative to the system with $N=5000$ particles
are quantitatively well-explained by Eq. (\ref{S_m_decay}) using
the values of $\eta$ from Fig. \ref{G4.domino.MC.power}. We note
that we have never observed a phase boundary between a crystallized
region and a disordered liquid, which would be indicative of a first-order
phase transition.


\subsection{\label{Section_S_k}Translational Order}

Measuring translational order is more difficult than orientational
ordering. From the observations above we are motivated to look for
translational order of the kind present in a random domino tiling.
Looking at the centroids of the dominos themselves does not reveal
a simple pattern. However, if we split each rectangle into two squares
and look at the centroids of the $2N$ squares, translational order
will be manifested through the appearance of square-lattice periodicity.
Such periodicity is most easily quantified by the Fourier transform
of the square centroids, i.e., the structure factor

\begin{equation}
S(\vec{k})=\frac{1}{N}\left|\sum_{j=1}^{N}\exp\left(i\V{k}\cdot\V{r}_{j}\right)\right|^{2}.\end{equation}
In a translationally disordered state, $S(\vec{k})$ is of order one
and decays to unity for large $k$. For long-ranged periodic systems,
$S(\vec{k})$ shows sharp Bragg peaks at the reciprocal lattice vectors,
while for quasi-long-range order the peaks have power-law wings. It
is however difficult to exactly determine when true peaks replace
the finite humps that exist due to short-range translational ordering
in the liquid state.

It would be convenient to have a scalar metric of translational order
similar to $S_{4}$ for tetratic order. We use the averaged value
of $S(\vec{k})$ over the first four Bragg peaks\begin{equation}
T_{k}=\frac{1}{2N}\left[S(\frac{2\pi}{\tilde{a}}\V{n}_{\Vert})+S(\frac{2\pi}{\tilde{a}}\V{n}_{\bot})\right],\label{S_k_scalar}\end{equation}
where $\tilde{a}=a/\sqrt{\phi}$ is the expected spacing of the underlying
square lattice, and $\V{n}_{\Vert}$ and $\V{n}_{\bot}$ are two perpendicular
unit vectors determining the orientation of the square lattice. In
the tiling limit $T_{k}=1$, and for a liquid $T_{k}\approx0$. When
decompressing a prepared tiling, we already know $\V{n}_{\Vert}=(1,0)$
and it is best to use this known value. However, when compressing
a liquid, we have no way of knowing the final orientation of the lattice
and therefore we use the bidirector $\V{n}_{\Vert}=\V{n}_{4}$, as
determined during the measurement of $S_{4}$. This method seems not
to work well because even small fluctuations in the director cause
large fluctuations in $T_{k}$, and additionally, small defects can
disrupt periodicity and significantly reduce the value of $T_{k}$
below unity. In Fig. \ref{S4.domino.hysteresis} we show the values
of $T_{k}$ along with $S_{4}$. It is seen that for the decompression
run, $T_{k}$ starts at unity and decays continuously until it apparently
goes to zero around $\phi\approx0.8$. The compression runs similarly
show a visible but noisy and masked increase above zero when the density
increases above $\phi\approx0.8$, but do not reach the same level
of $T_{k}$ as for decompression from a perfect crystal. We are therefore
led to believe that there is a second transition from tetratic liquid
to tetratic solid at $\phi\approx0.8$.

\begin{figure}
\begin{center}\includegraphics[%
  width=0.7\textwidth,
  keepaspectratio]{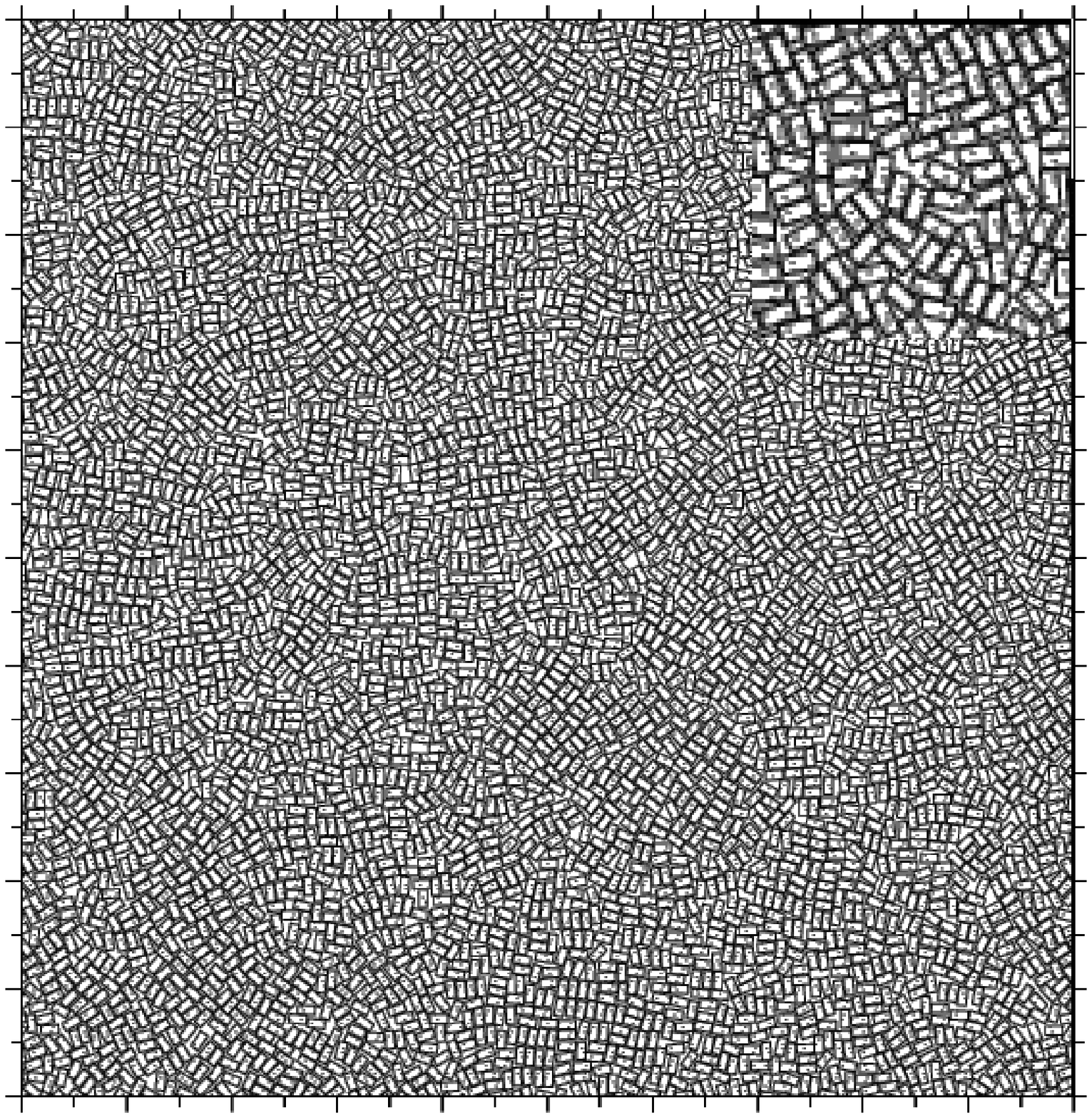}\end{center}

\begin{center}\includegraphics[%
  width=0.45\textwidth,
  keepaspectratio]{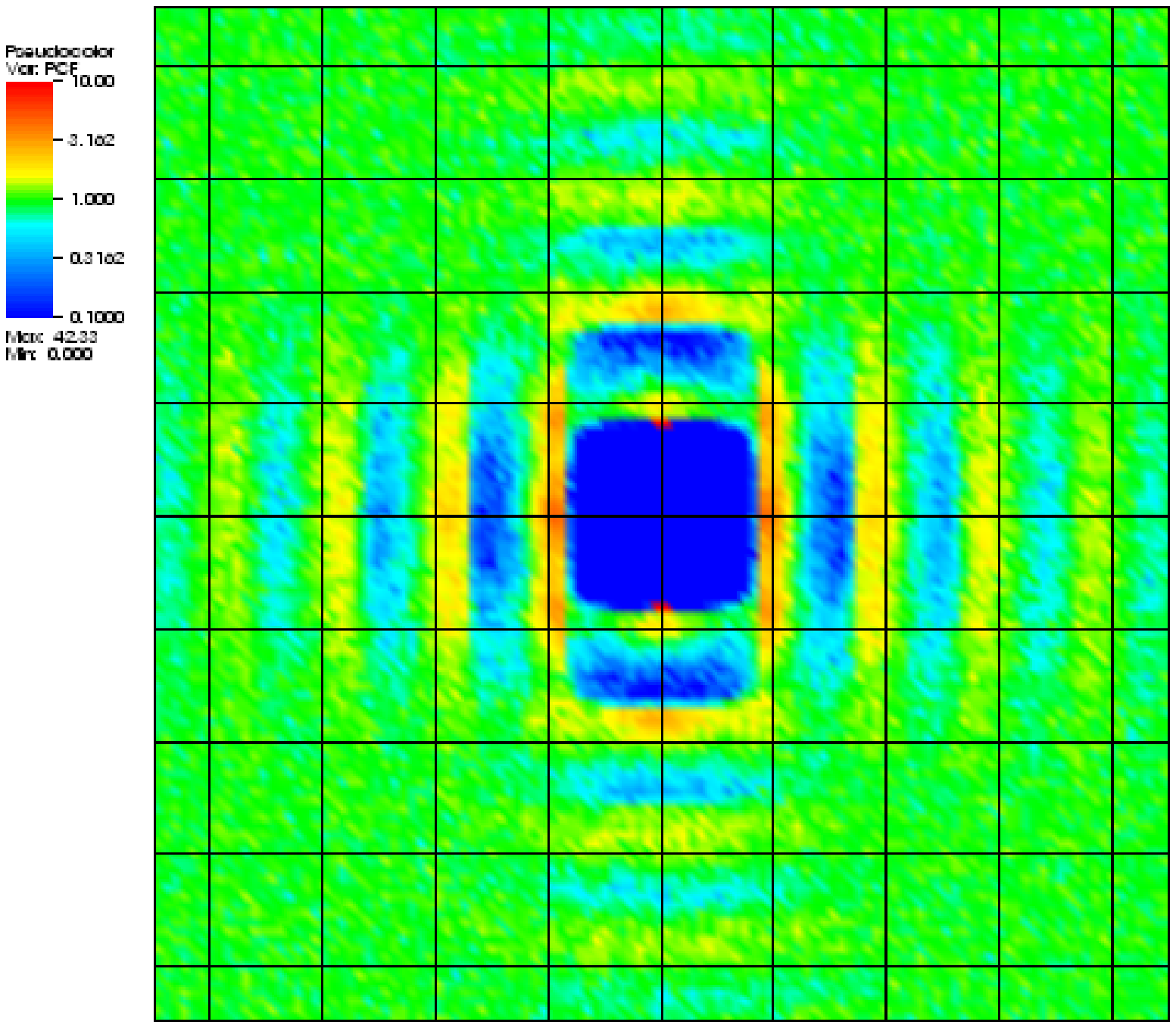}\includegraphics[%
  width=0.45\textwidth,
  keepaspectratio]{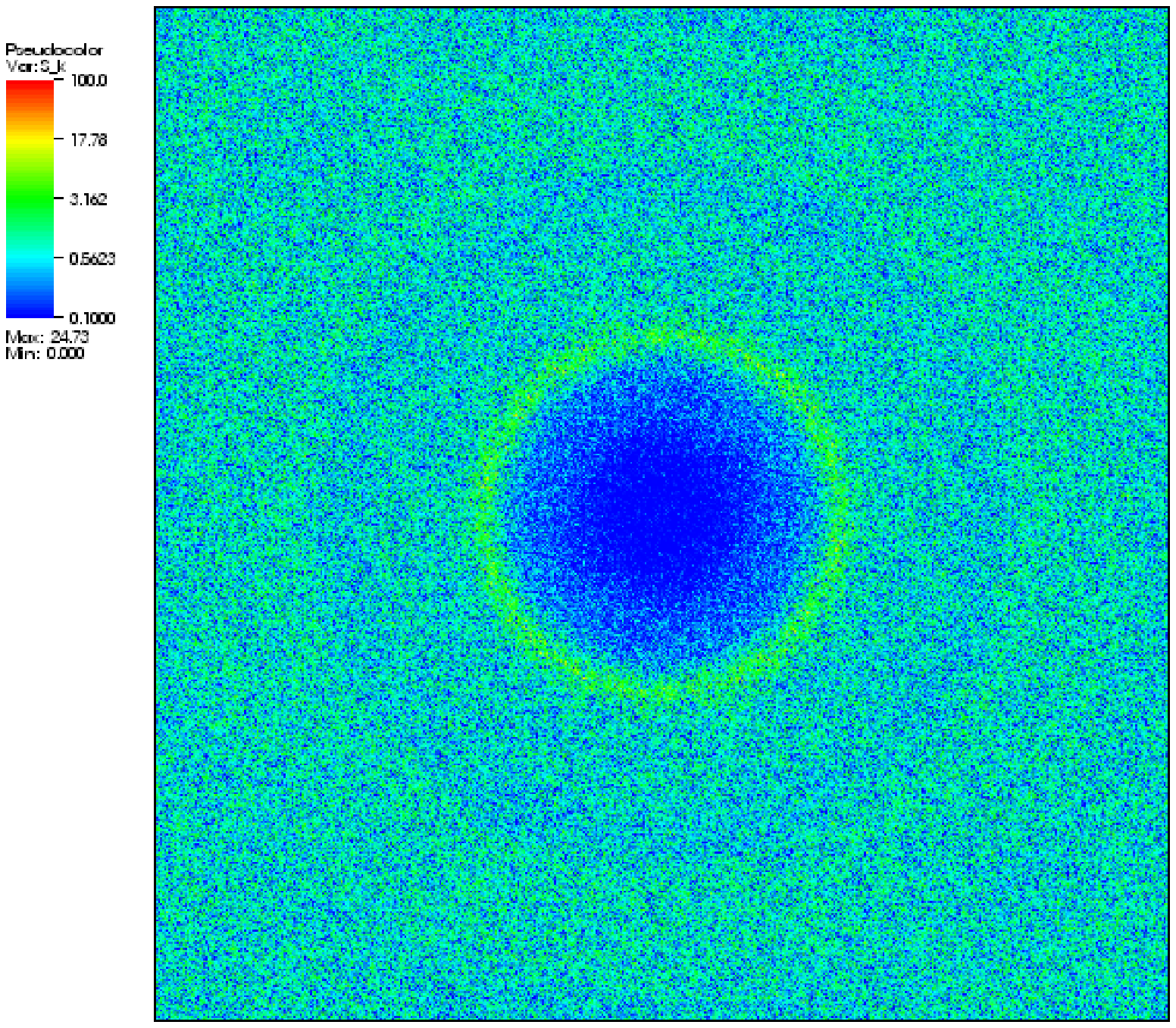}\end{center}

\caption{\label{Domino.diced.phi=3D0.700}(Color online) A snapshot configuration
of a system of $N=5000$ dominos at $\phi=0.7$ (top) with inset with
threefold magnification showing local packing structure, along with
$g_{2}(\D{x},\D{y})$ overlayed over the underlying square lattice
(bottom left) and $S(\vec{k})$ (bottom right), obtained after splitting
each domino into two squares. It is clear that the system is isotropic
from the rotational symmetry of $S(\vec{k})$. Only short-range order
is visible in $g_{2}(\D{x},\D{y})$, confirming that this is an isotropic
liquid.}
\end{figure}

\begin{figure}
\begin{center}\includegraphics[%
  width=0.7\textwidth,
  keepaspectratio]{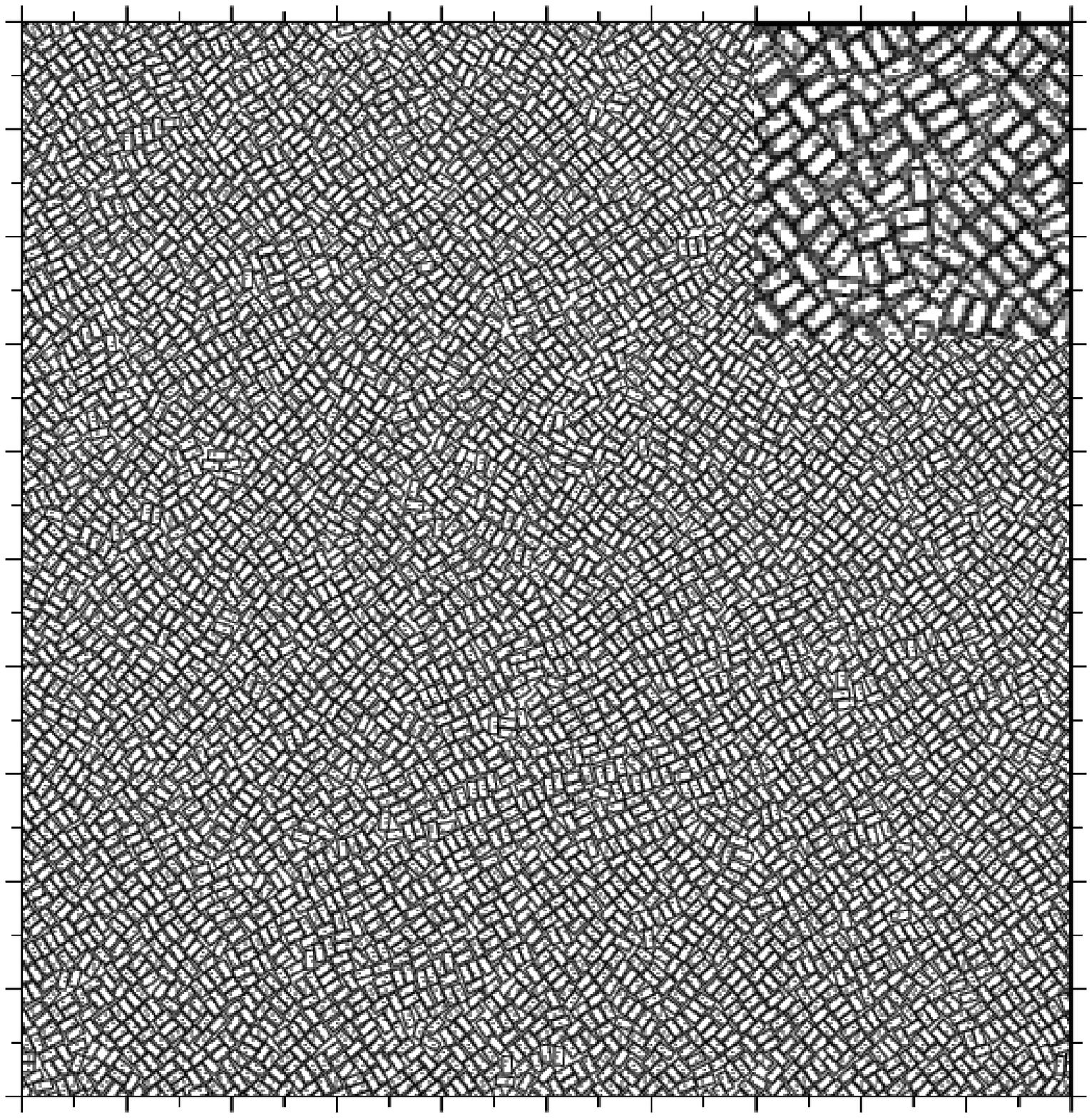}\end{center}

\begin{center}\includegraphics[%
  width=0.45\textwidth,
  keepaspectratio]{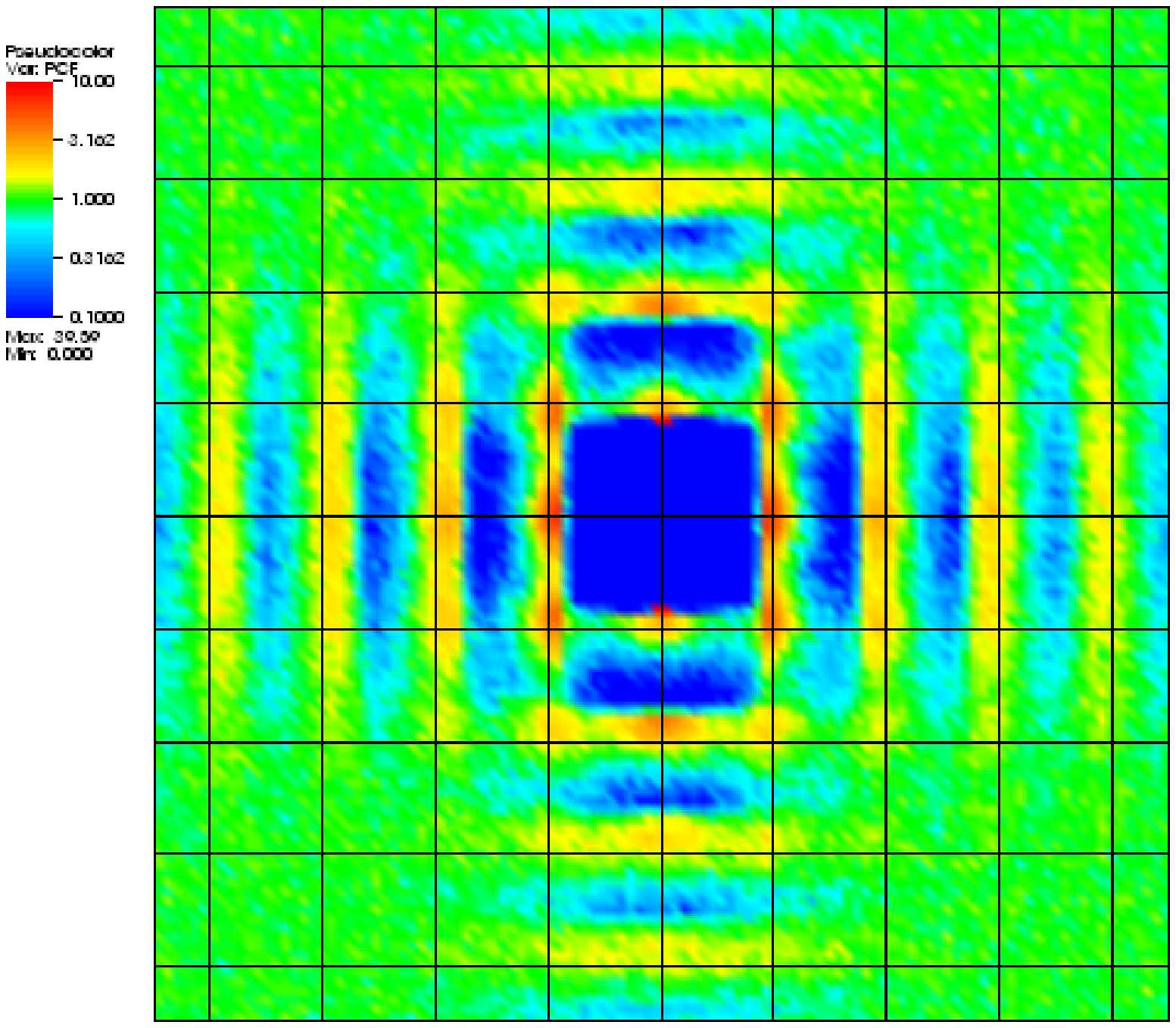}\includegraphics[%
  width=0.45\textwidth,
  keepaspectratio]{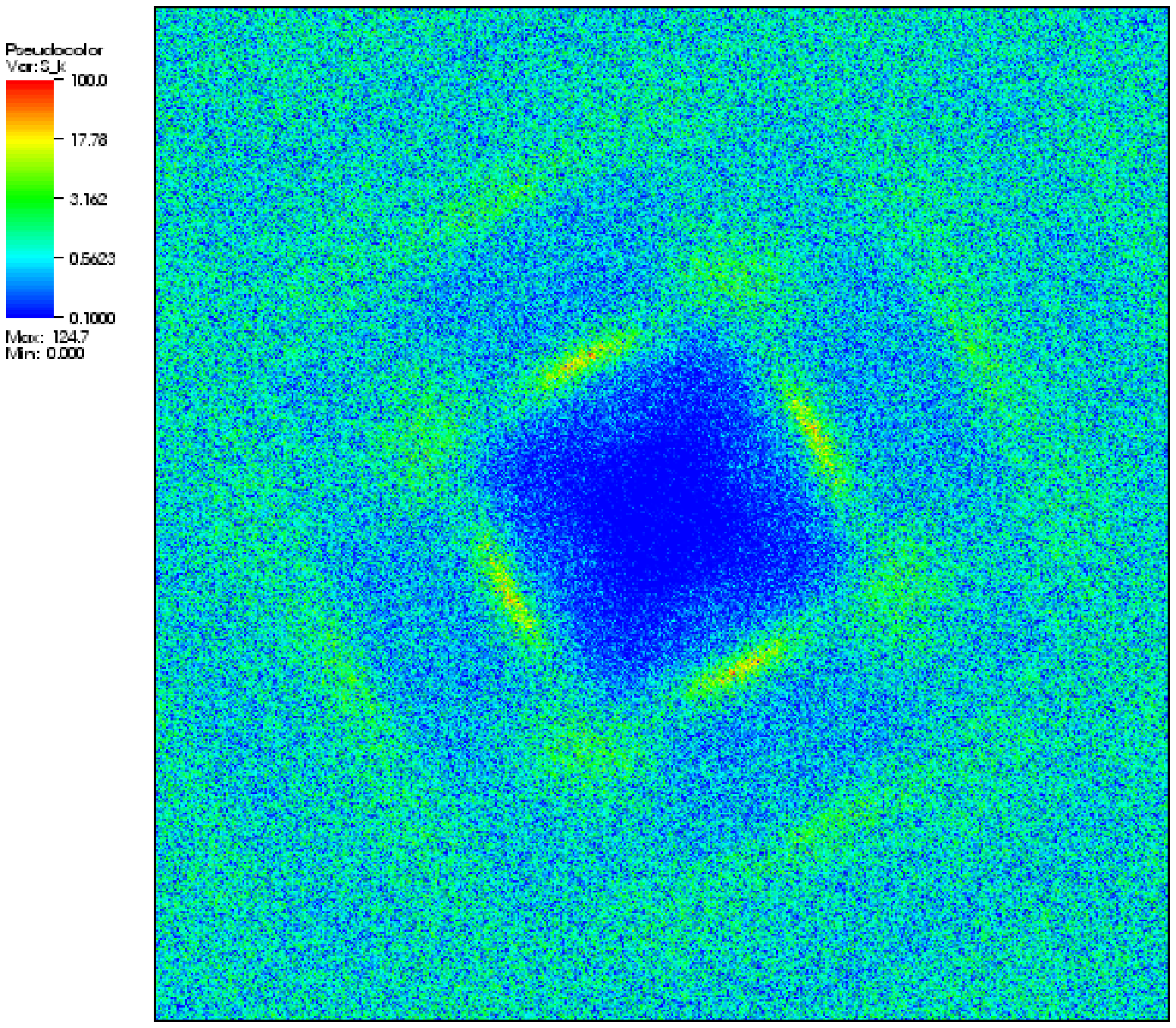}\end{center}

\caption{\label{Domino.diced.phi=3D0.750}(Color online) A system of $N=5000$
dominos as in Fig. \ref{Domino.diced.phi=3D0.700} but at $\phi=0.750$,
which shows a tetratic liquid phase. Four-fold broken symmetry is
seen in $S(\vec{k})$, but without pronounced sharp peaks. The range
of ordering in $g_{2}(r)$ has increased, but still appears of much
shorter range than the size of the system, as seen clearly in the
plot of the actual domino configuration. It is interesting that $g_{2}(\D{x},\D{y})$
is very anisotropic, being much stronger to the side of a square relative
to its diagonals. No phase boundary characteristic of first-order
transitions is visible.}
\end{figure}

\begin{figure}
\begin{center}\includegraphics[%
  width=0.7\textwidth,
  keepaspectratio]{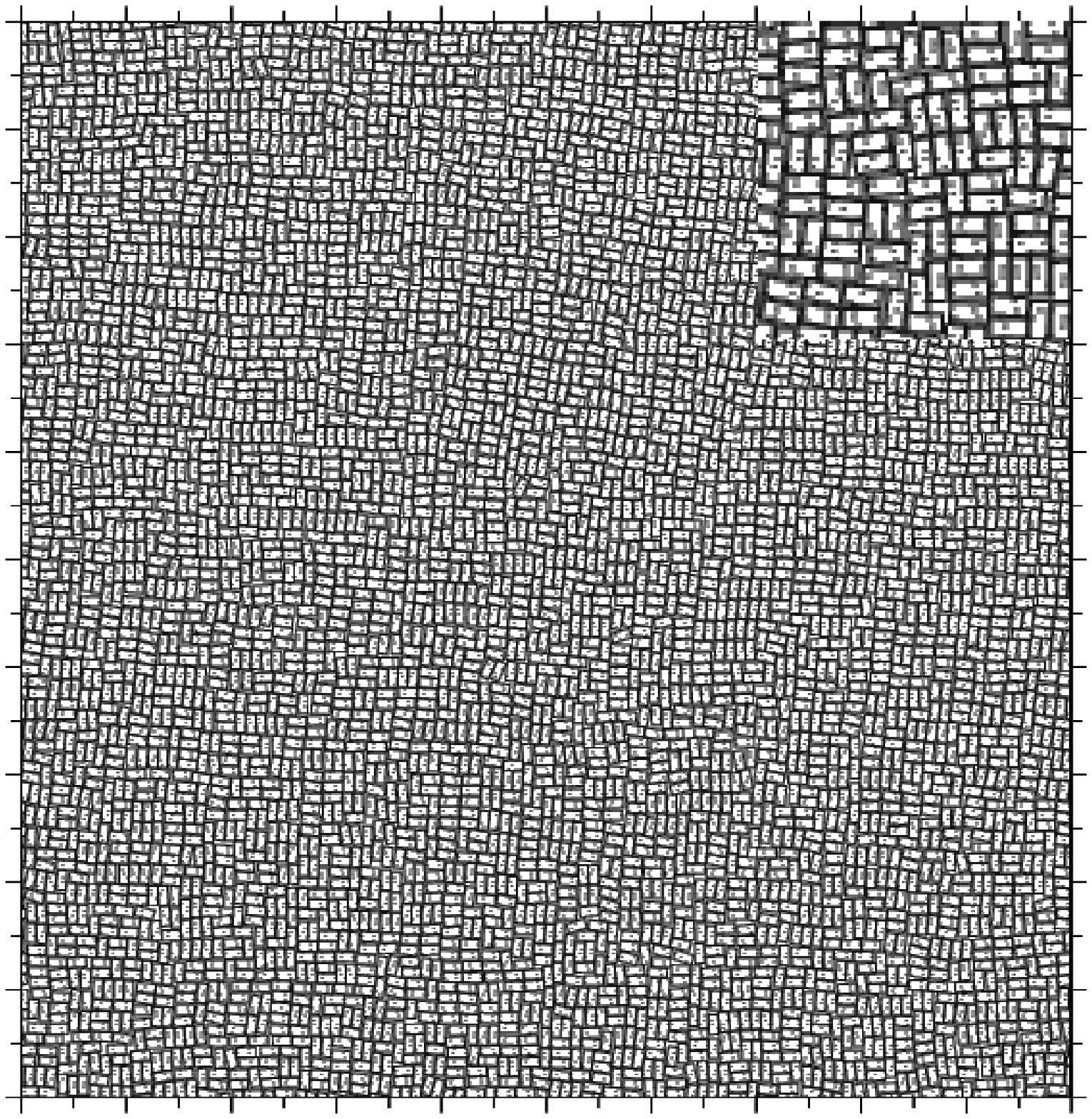}\end{center}

\begin{center}\includegraphics[%
  width=0.45\textwidth,
  keepaspectratio]{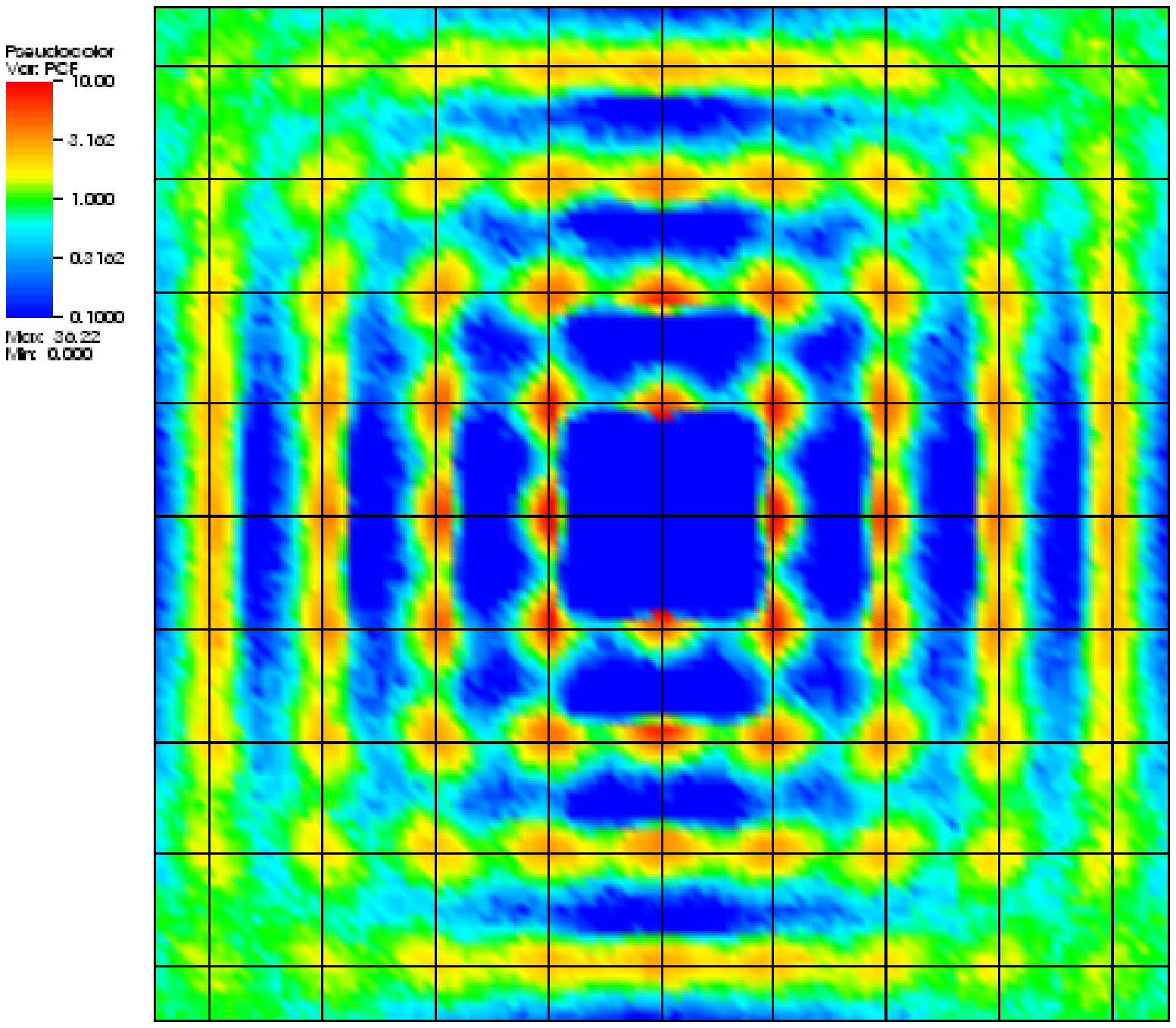}\includegraphics[%
  width=0.45\textwidth,
  keepaspectratio]{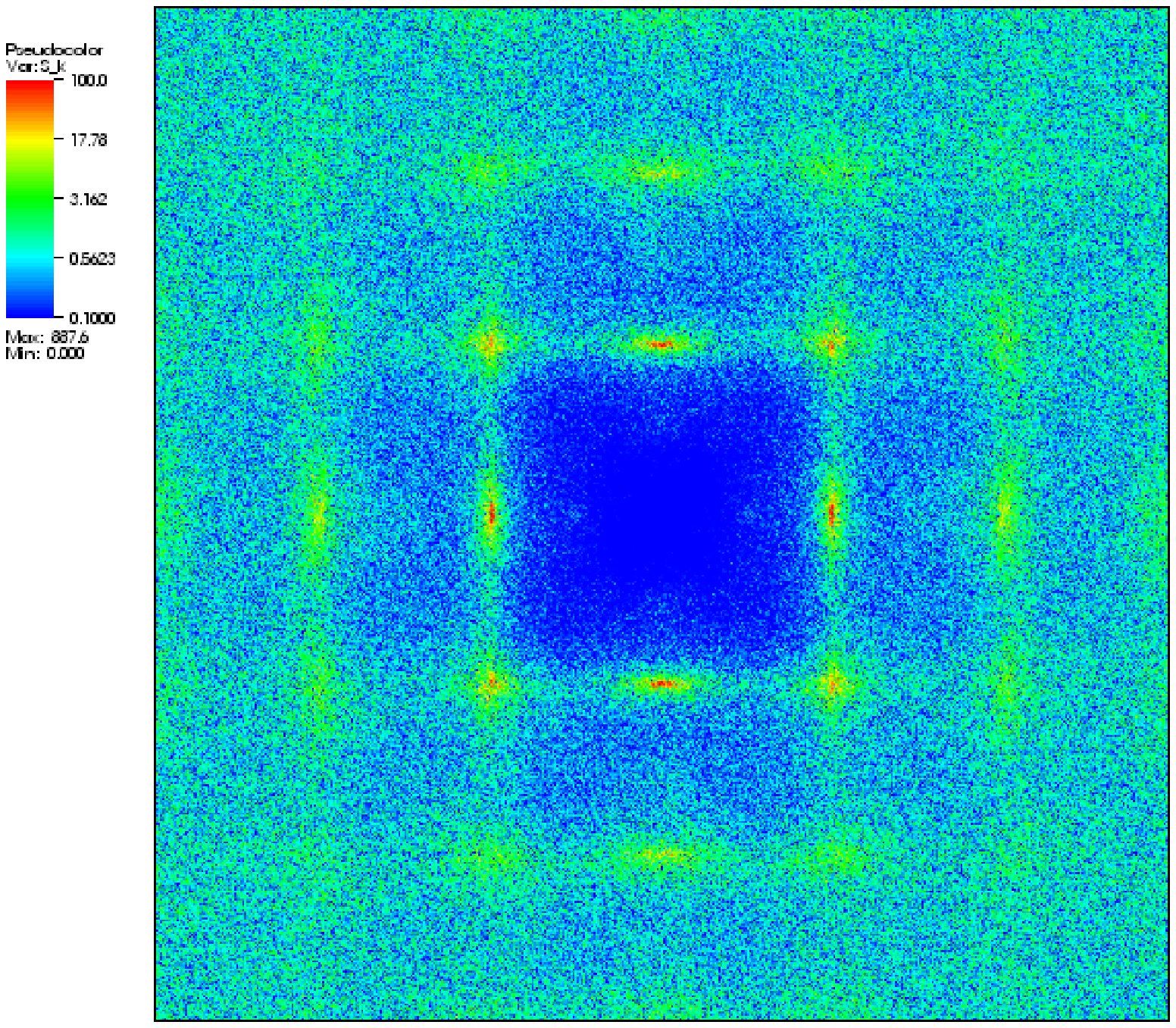}\end{center}

\caption{\label{Domino.diced.phi=3D0.825}(Color online) A system of $N=5000$
dominos as in Figs. \ref{Domino.diced.phi=3D0.700} and \ref{Domino.diced.phi=3D0.750}
but at $\phi=0.825$. The structure factor shows sharp peaks (maximum
value is above $10$) on the sites of a (reciprocal) square lattice,
and $g_{2}(r)$ shows longer-ranged translational ordering, indicating
a solid phase. Visual inspection of the configuration confirms that
the translational ordering spans the system size and shows some vacancies
consisting of only a single square (half a particle).}
\end{figure}

In addition to reciprocal space $S(\vec{k})$, one can also look at
the center-to-center-distance distribution function $g_{2}(r)$ for
the squares (half dominos). However, quantitative analysis of $g_{2}(r)$
is made difficult because of oscillations due to exclusion effects
and also due to the coupling to orientation. Instead of presenting
such a one-dimensional pair correlation function, we present $g_{2}(\D{x},\D{y})$,
which is simply the orientationally-averaged $g_{2}(\D{x},\D{y},\D{\theta})$.
In Figs. \ref{Domino.diced.phi=3D0.700}, \ref{Domino.diced.phi=3D0.750}
and \ref{Domino.diced.phi=3D0.825} we show a snapshot of a system
of $N=5000$ dominos, along with the corresponding $g_{2}(\D{x},\D{y})$
and $S(\vec{k})$, for three densities, corresponding to an isotropic
liquid, a tetratic liquid {[}i.e., a state with (quasi-) long range
tetratic but only short-range translational order{]}, and a tetratic
solid {[}i.e., a state with (quasi-) long range tetratic and translational
order{]}. For the $g_{2}(\D{x},\D{y})$ plots, we have drawn the expected
underlying square lattice at that density. Note that $g_{2}(\D{x},\D{y})$
always has two sharp peaks corresponding to the square glued to the
one under consideration in the dimer (domino).

\subsection{\label{Section_Tiling}Solid Phase}

Having confirmed the appearance of a translationally-ordered tetratic
phase closely related to domino tilings of the plane, we turn to understanding
the nature of the the thermodynamically-favored tiling. There are
two likely possibilities: The tiling shows (translational) ordering
itself, or the tiling is {}``random''. In the context of a discrete
system like domino tilings, the concept of a random tiling is mathematically
well-defined in terms of maximizing entropy\periodcite{RandomTilings_Entropy,RandomTilings_Math}
This random tiling has a positive degeneracy entropy $0.58313k_{B}$,
unlike ordered tilings such as the nematic tiling (in which all dominos
are alligned).

\begin{figure}
\begin{center}\includegraphics[%
  width=0.95\columnwidth,
  keepaspectratio]{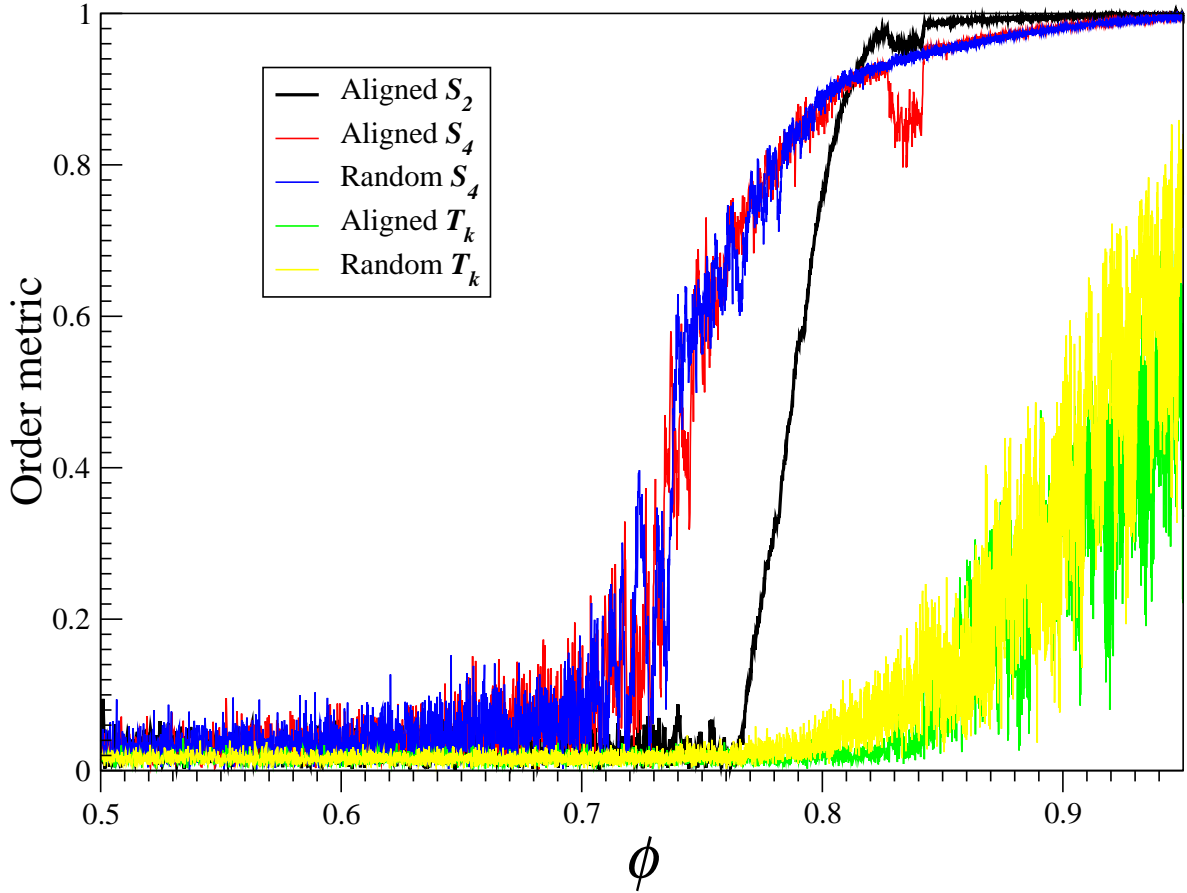}\end{center}

\caption{\label{S4.nematic.MD}Nematic, tetratic, and translational order
metrics as a domino tiling in which all rectangles are aligned is
slowly decompressed from close-packing. The nematic crystal spontaneously
realigned to a different orientation of the director from the starting
one at around $\phi\approx0.84$, causing some fluctuations and a
drop in $T_{k}$ which are likely just a finite-size (boundary) effect.}
\end{figure}

Our compressions of isotropic liquids have invariably led to apparently
disordered domino tilings upon spontaneous {}``freezing'', albeit
with some frozen defects. This suggests that the disordered tiling
has lower free energy than ordered tilings. However, it is also possible
that the disorder is simply dynamically trapped when the tetratic
liquid freezes. In fact, starting a decompression run from an aligned
nematic tiling shows that the tiling configuration is preserved until
melting into a tetratic liquid occurs around $\phi\approx0.8$. This
is demonstrated in Fig. \ref{S4.nematic.MD}, where both $S_{4}$
and $S_{2}$ as well as $T_{k}$ are shown along a decompression run
starting with both a disordered and an ordered tiling. It is seen
that $S_{2}$ drops sharply around $\phi\approx0.8$ while $S_{4}$
remains positive until $\phi\approx0.7$, clearly demonstrating the
thermodynamic stability of the tetratic liquid phase in the intermediate
density range. Subsequent compression of this liquid would lead to
a disordered tiling without any trace of the initial nematic ordering.

We expect that the free-volume contribution to the free energy is
minimized for ordered tilings at high densities. However, we also
expect that solid phase is ergodic in the sense that transitions between
alternative tiling configurations will occur in long runs of very
large systems, so that in the thermodynamic limit the space of all
tilings will be explored. This amounts to a positive contribution
to the entropy of the disordered tiling due to its degeneracy, and
it is this entropy that can thermodynamically stabilize the disordered
tiling even in the close-packed limit. A closer analysis similar to
that carried for hard-disk dimers in Refs. \onlinecite{DiskDimers_Solid,DiskDimers_DC}
is necessary. In particular, including collective Monte Carlo trial
moves that transition between different tiling configurations, as
well as relaxation of the dimensions of the unit cell (important for
smaller solid systems), is important. We believe that, just like the
hard-disk dimer system, the hard-square dimer system has a thermodynamically
stable nonperiodic solid phase. Also, as in the hard dumbbell (fused
hard-disk dimers) system, we expect that for aspect ratios close to,
but not exactly, two, the nonperiodic solid will be replaced by a
nematic (and possibly periodic) phase at the highest densities\periodcite{DiskDimers_DC,DiskDimers_FreeVolume}
This is because reaching the maximal density $\phi=1$ seems to require
aligning the rectangles. It is interesting, however, that at least
for rational, and certainly for integer aspect rations such as $\alpha=3$,
there is the possibility of disordered solid phases being stable even
in the close-packed limit. On physical grounds we expect the phase
diagram to vary smoothly with aspect ratio, rather than depending
sensitively on the exact value of $\alpha$.

\begin{figure}
\begin{center}\subfigure{\label{SHE.domino.N=3D5000.random}\includegraphics[%
  width=0.35\textheight,
  keepaspectratio]{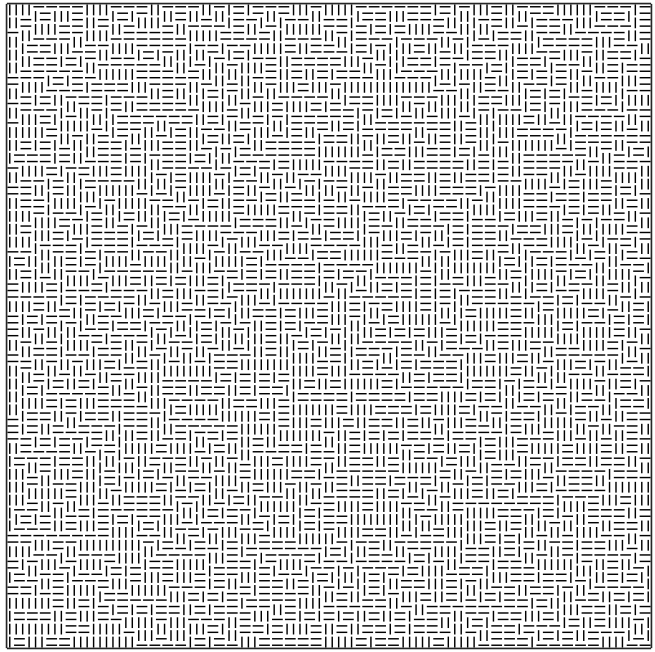}}\end{center}

\begin{center}\subfigure{\label{LSD.domino.N=3D5000.solid}\includegraphics[%
  width=0.35\textheight,
  keepaspectratio]{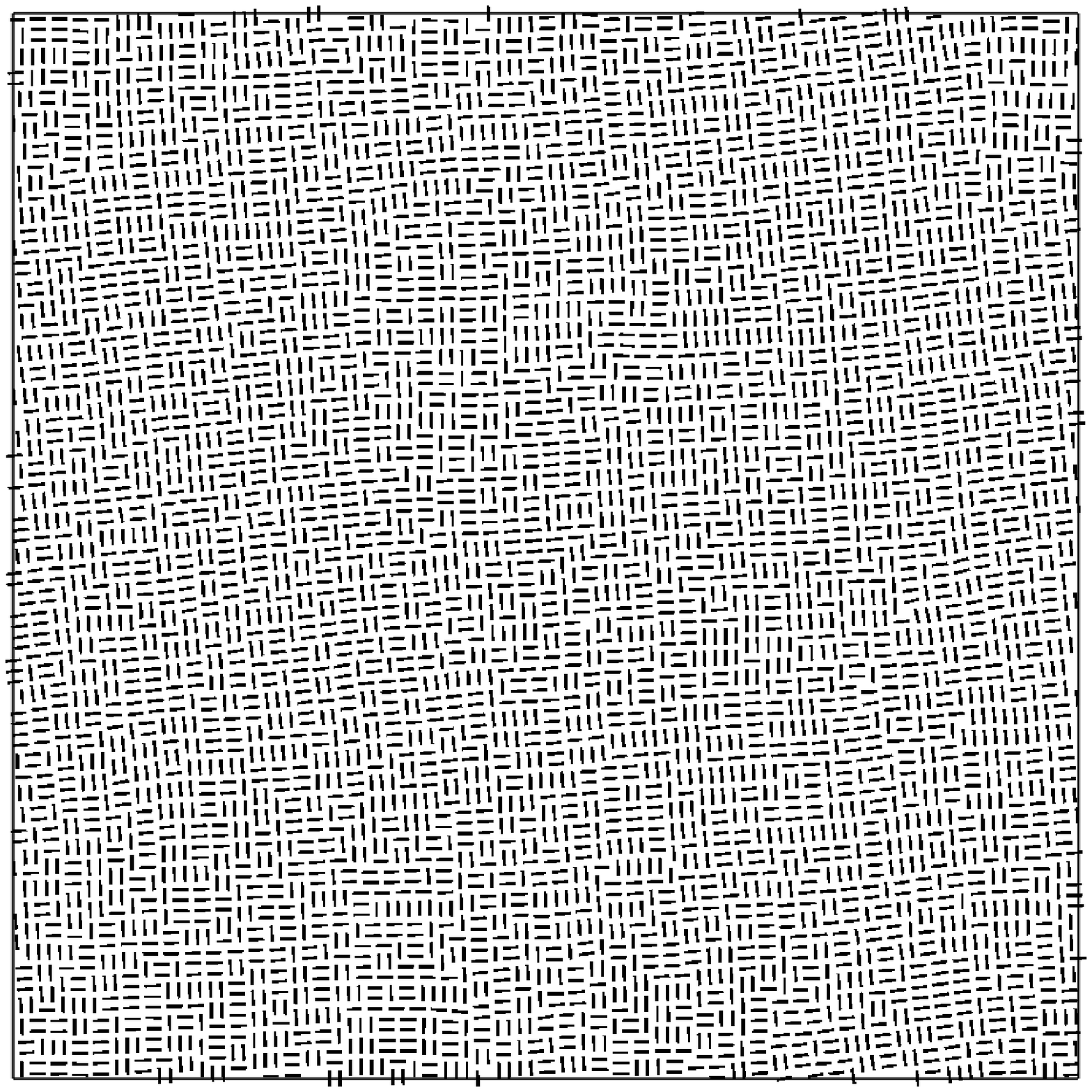}}\end{center}

\caption{\label{CompareDomino_SHE} A comparison between a true random tiling
of a square with dominos\spacecite{RandomTiling_SpanningTree} (top),
and the unit cell of a system of $N=5000$ superellipses with exponent
$\zeta=7.5$ slowly compressed from isotropic liquid to $\phi=0.95$
(bottom). The compressed system is not a perfect tiling due its lower
density and frozen defects, as well as the rounding of the superellipses
relative to true rectangles. Therefore at large scales the two systems
look different. However a closer local examination reveals similar
tiling patterns in the two systems, typical of {}``random'' tilings.}
\end{figure}

\begin{figure}
\begin{center}\includegraphics[%
  width=0.5\linewidth,
  keepaspectratio]{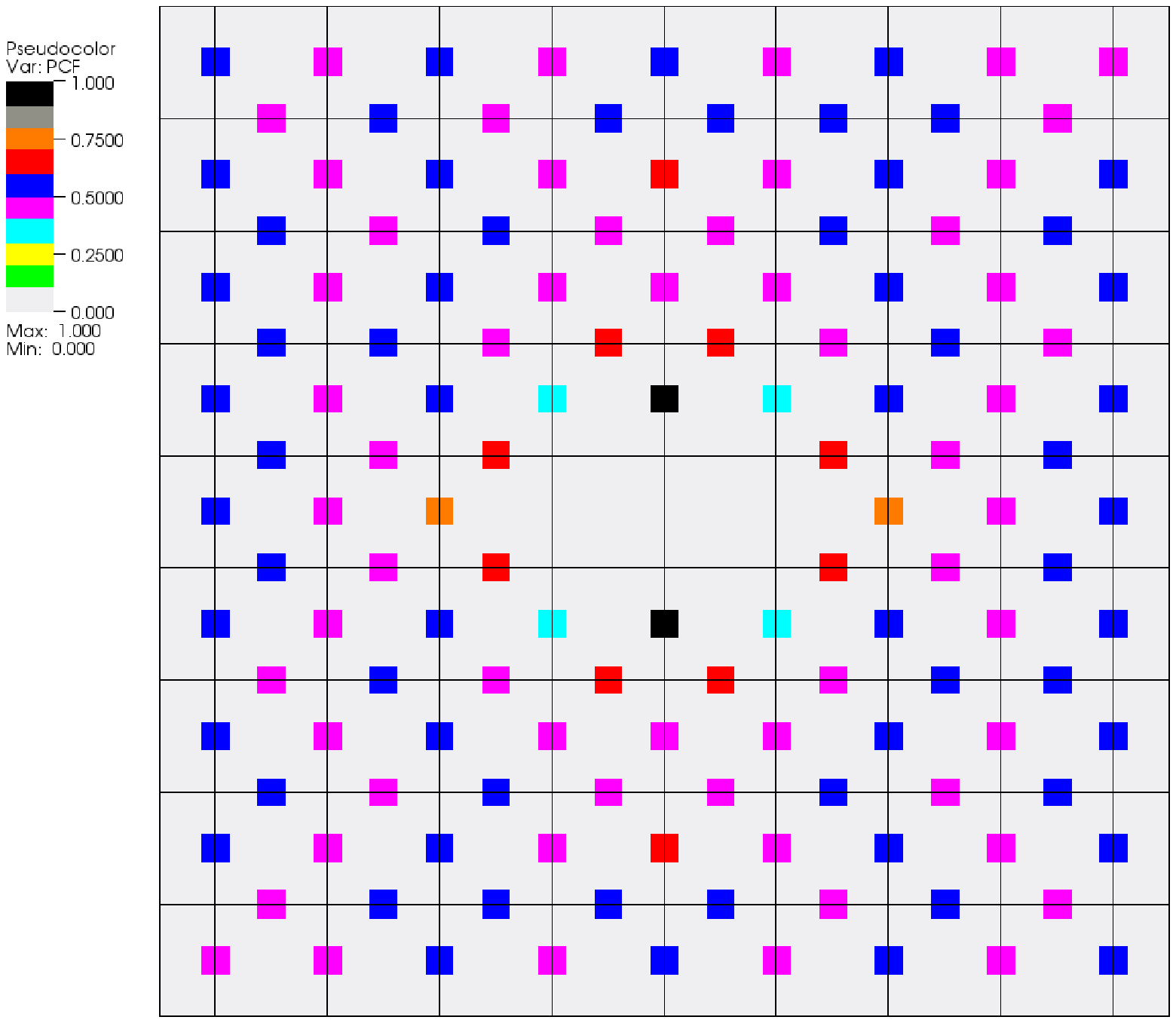}\includegraphics[%
  width=0.5\linewidth,
  keepaspectratio]{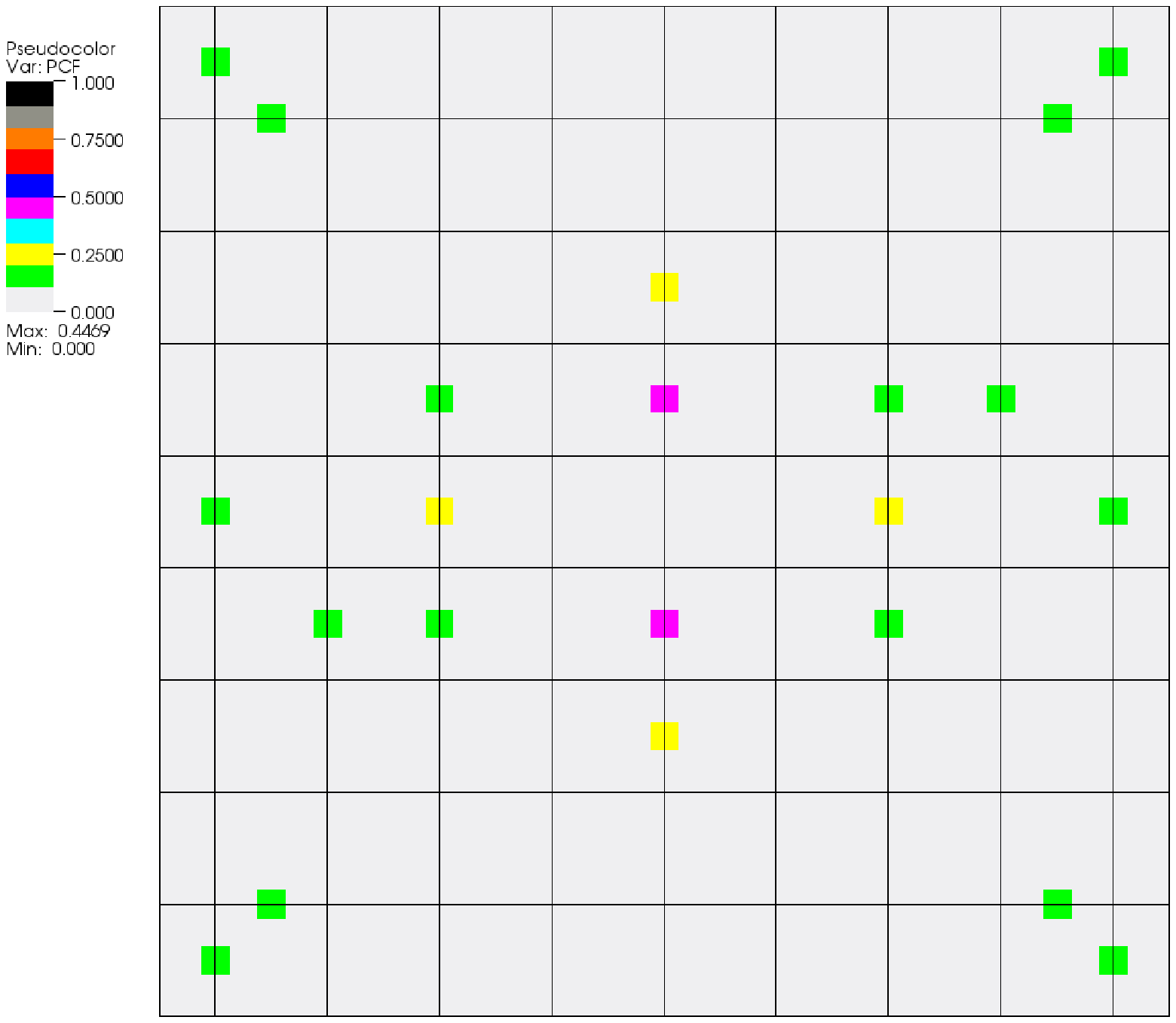}\end{center}

\caption{\label{Dominos_PCF}(Color online) \emph{Left}: Center-center pair
correlation function $g_{2}(\D{x},\D{y})$ for the perfect random
tiling in Fig. \ref{CompareDomino_SHE}. This $g_{2}$ is a collection
of $\delta$-functions whose heights can also be calculated exactly\spacecite{DimerCovering_Statistics} 
(the calculation is nontrivial and
we have not performed it). We have normalized $g_{2}$ so that the
highest peaks have a value of one. \emph{Right}: The absolute value
of the difference between $g_{2}(\D{x},\D{y})$ for the two systems
shown in Fig. \ref{CompareDomino_SHE}, shown on a coarse-enough scale
so that the broadening of the peaks due to thermal motion is not visible.
The color table used in this figure is discrete in order to highlight
the symmetry and hide small fluctuations due to finite system size.
The difference in $g_{2}$ is almost entirely within the smallest
interval of the color table (less than $0.1$, gray), with only some
peaks showing differences up to $0.25$.}
\end{figure}

Accepting for a moment the existence of a nonperiodic solid phase,
it remains to verify that the compressed systems we obtain in our
simulations are indeed like (maximal entropy) random tilings of the
plane with dimers. This is hard to do rigorously, as it requires comparing
all correlation functions between a random tiling and our compressed
systems. Figure \ref{CompareDomino_SHE} shows a visual comparison
of a random tiling of a large square, generated using random spanning
trees by a program provided to us by the authors of Ref. \onlinecite{RandomTiling_SpanningTree},
and a system of superellipses compressed to $\phi=0.95$ (close to
the achievable maximum for our MD program for such high superellipse
exponents). While the translational ordering in the compressed solid
is clearly not perfect as it is for the true tiling, visual inspection
suggests close similarity between the the local tiling patterns of
the two systems. In Fig. \ref{Dominos_PCF}, we show $g_{2}(\D{x},\D{y})$
for the true tiling, along with the difference in $g_{2}$ between
the true tiling and the compressed solid. Here we do not split the
rectangles into two squares, i.e., the figure shows the probability
density of observing a centroid of another rectangle at $(\D{x},\D{y})$
given a rectangle at the origin oriented with the long side along
the $x$ axis. It can be seen that there is a close match between
the random tiling and the compressed solid, at least at the two-body
correlation level.

\section{\label{Section_Conclusions}Conclusions and Future Directions}

The results presented in this paper highlight the unusual properties
of the simple hard-rectangle system when the aspect ratio is $\alpha=2$,
hopefully stimulating further research into the hard-rectangle system.
For square dimers (dominos), in addition to the expected low-density
isotropic liquid phase, a stable tetratic liquid phase is clearly
observed, in which there is four-fold orientational ordering but no
translational ordering. A tetratic solid phase closely connected to
random domino tilings is found and we conjecture that it is thermodynamically
stabilized by its positive degeneracy entropy. The transitions between
the phases are consistent with a KTHNY-like sequence of two continuous
transitions. If this is indeed the case, then the hard dimer system
provides an excellent model for the study of continuous transitions,
with a rather large gap in density between the two presumed transitions
$\D{\phi}\approx0.1$, unlike the hard-disk system. Random jammed
packings of rectangles seem to be translationally ordered, similar
to the behavior for disks\spacecite{Jamming_LP_results} but unlike spheres
which can jam in disordered configurations\periodcite{Torquato_MRJ} However,
unlike disks, the systems of rectangles show orientational disorder,
once again illustrating the geometric richness of even the simplest
hard-particle models.

Further investigations are needed for the domino system to conclusively
determine its phase behavior. Improved MC with collective moves that
explore multiple tilings, as well as allow for relaxation of the boundary
conditions, should be implemented. Additionally, the free energies
of the different phases should be computed so that the exact locations
of the phase transitions could be identified. The final goal is to
completely characterize the phase diagram of the hard rectangle system
in the $\alpha-\phi$ plane, as has been done, for example, for 
diskorectangles\spacecite{SpherocyllindersPhases_2D} and ellipses\periodcite{EllipsesPhases_MC}
In addition to nematic and smectic phases, novel liquid crystal phases
with tetratic order may be discovered.

\begin{acknowledgments}
The authors were supported in part by the National Science Foundation
under Grant No. DMS-0312067. We thank Paul Chaikin for stimulating
our interest in this problem and for numerous helpful discussions.
We thank David Wilson for providing us with a program to generate
random domino tilings\periodcite{RandomTiling_SpanningTree}
\end{acknowledgments}

\begin{thebibliography}{30}
\expandafter\ifx\csname natexlab\endcsname\relax\def\natexlab#1{#1}\fi
\expandafter\ifx\csname bibnamefont\endcsname\relax
  \def\bibnamefont#1{#1}\fi
\expandafter\ifx\csname bibfnamefont\endcsname\relax
  \def\bibfnamefont#1{#1}\fi
\expandafter\ifx\csname citenamefont\endcsname\relax
  \def\citenamefont#1{#1}\fi
\expandafter\ifx\csname url\endcsname\relax
  \def\url#1{\texttt{#1}}\fi
\expandafter\ifx\csname urlprefix\endcsname\relax\def\urlprefix{URL }\fi
\providecommand{\bibinfo}[2]{#2}
\providecommand{\eprint}[2][]{\url{#2}}

\bibitem[{\citenamefont{Barrat and Hansen}(2003)}]{SimpleComplex_Liquids}
\bibinfo{author}{\bibfnamefont{J.-L.} \bibnamefont{Barrat}} \bibnamefont{and}
  \bibinfo{author}{\bibfnamefont{J.-P.} \bibnamefont{Hansen}},
  \emph{\bibinfo{title}{{Basic Concepts for Simple and Complex Liquids}}}
  (\bibinfo{publisher}{Cambridge University Press}, \bibinfo{year}{2003}).

\bibitem[{\citenamefont{Camp and Allen}(1997)}]{Biaxial_Ellipsoid_Fluid}
\bibinfo{author}{\bibfnamefont{P.~J.} \bibnamefont{Camp}} \bibnamefont{and}
  \bibinfo{author}{\bibfnamefont{M.~P.} \bibnamefont{Allen}},
  \bibinfo{journal}{J. Chem. Phys.} \textbf{\bibinfo{volume}{106}},
  \bibinfo{pages}{6681} (\bibinfo{year}{1997}).

\bibitem[{\citenamefont{Acharya et~al.}(2004)\citenamefont{Acharya, Primak, and
  Kumar}}]{BiaxialNematic_Boomerang}
\bibinfo{author}{\bibfnamefont{B.~R.} \bibnamefont{Acharya}},
  \bibinfo{author}{\bibfnamefont{A.}~\bibnamefont{Primak}}, \bibnamefont{and}
  \bibinfo{author}{\bibfnamefont{S.}~\bibnamefont{Kumar}},
  \bibinfo{journal}{Phys. Rev. Lett.} \textbf{\bibinfo{volume}{92}},
  \bibinfo{pages}{145506} (\bibinfo{year}{2004}).

\bibitem[{\citenamefont{Wojciechowski and Frenkel}(2004)}]{HardSquares_MC}
\bibinfo{author}{\bibfnamefont{K.~W.} \bibnamefont{Wojciechowski}}
  \bibnamefont{and} \bibinfo{author}{\bibfnamefont{D.}~\bibnamefont{Frenkel}},
  \bibinfo{journal}{Comp. Methods in Science and Tech.}
  \textbf{\bibinfo{volume}{10}}, \bibinfo{pages}{235} (\bibinfo{year}{2004}).

\bibitem[{\citenamefont{Wojciechowski et~al.}(1991)\citenamefont{Wojciechowski,
  Frenkel, and Bra\'{n}ka}}]{DiskDimers_Solid}
\bibinfo{author}{\bibfnamefont{K.~W.} \bibnamefont{Wojciechowski}},
  \bibinfo{author}{\bibfnamefont{D.}~\bibnamefont{Frenkel}}, \bibnamefont{and}
  \bibinfo{author}{\bibfnamefont{A.}~\bibnamefont{Bra\'{n}ka}},
  \bibinfo{journal}{Phys. Rev. Lett.} \textbf{\bibinfo{volume}{66}},
  \bibinfo{pages}{3168} (\bibinfo{year}{1991}).

\bibitem[{\citenamefont{Mart\'{i}nez-Rat\'{o}n
  et~al.}(2005)\citenamefont{Mart\'{i}nez-Rat\'{o}n, Velasco, and
  Mederos}}]{HardRodFluids_DFT}
\bibinfo{author}{\bibfnamefont{Y.}~\bibnamefont{Mart\'{i}nez-Rat\'{o}n}},
  \bibinfo{author}{\bibfnamefont{E.}~\bibnamefont{Velasco}}, \bibnamefont{and}
  \bibinfo{author}{\bibfnamefont{L.}~\bibnamefont{Mederos}},
  \bibinfo{journal}{J. of Chem. Phys.} \textbf{\bibinfo{volume}{122}}
  (\bibinfo{year}{2005}).

\bibitem[{\citenamefont{Schlacken et~al.}(1998)\citenamefont{Schlacken, Mogel,
  and Schiller}}]{HardRodFluids_SPT}
\bibinfo{author}{\bibfnamefont{H.}~\bibnamefont{Schlacken}},
  \bibinfo{author}{\bibfnamefont{H.-J.} \bibnamefont{Mogel}}, \bibnamefont{and}
  \bibinfo{author}{\bibfnamefont{P.}~\bibnamefont{Schiller}},
  \bibinfo{journal}{Mol. Phys.} \textbf{\bibinfo{volume}{93}},
  \bibinfo{pages}{777} (\bibinfo{year}{1998}).

\bibitem[{\citenamefont{Kasteleyn}(1961)}]{DimerCovering_square_1}
\bibinfo{author}{\bibfnamefont{P.~W.} \bibnamefont{Kasteleyn}},
  \bibinfo{journal}{Physica} \textbf{\bibinfo{volume}{27}},
  \bibinfo{pages}{1209} (\bibinfo{year}{1961}).

\bibitem[{\citenamefont{Temperley and Fisher}(1961)}]{DimerCovering_square_2}
\bibinfo{author}{\bibfnamefont{H.~N.~V.} \bibnamefont{Temperley}}
  \bibnamefont{and} \bibinfo{author}{\bibfnamefont{M.~E.}
  \bibnamefont{Fisher}}, \bibinfo{journal}{Phil. Mag.}
  \textbf{\bibinfo{volume}{6}}, \bibinfo{pages}{1061} (\bibinfo{year}{1961}).

\bibitem[{\citenamefont{Kosterlitz and Thouless}(1973)}]{KT_Transition_2D}
\bibinfo{author}{\bibfnamefont{J.~M.} \bibnamefont{Kosterlitz}}
  \bibnamefont{and} \bibinfo{author}{\bibfnamefont{D.~J.}
  \bibnamefont{Thouless}}, \bibinfo{journal}{J. Phys.}
  \textbf{\bibinfo{volume}{C6}}, \bibinfo{pages}{1181} (\bibinfo{year}{1973}).

\bibitem[{\citenamefont{Halperin and Nelson}(1979)}]{HN_Transition_2D}
\bibinfo{author}{\bibfnamefont{B.~I.} \bibnamefont{Halperin}} \bibnamefont{and}
  \bibinfo{author}{\bibfnamefont{D.~R.} \bibnamefont{Nelson}},
  \bibinfo{journal}{Phys. Rev. B} \textbf{\bibinfo{volume}{19}},
  \bibinfo{pages}{2457} (\bibinfo{year}{1979}).

\bibitem[{\citenamefont{Young}(1979)}]{Young_Transition_2D}
\bibinfo{author}{\bibfnamefont{A.~P.} \bibnamefont{Young}},
  \bibinfo{journal}{Phys. Rev. B} \textbf{\bibinfo{volume}{19}},
  \bibinfo{pages}{1855} (\bibinfo{year}{1979}).

\bibitem[{\citenamefont{Weber and Stillinger}(1993)}]{SquareCrystal_3body}
\bibinfo{author}{\bibfnamefont{T.~A.} \bibnamefont{Weber}} \bibnamefont{and}
  \bibinfo{author}{\bibfnamefont{F.~H.} \bibnamefont{Stillinger}},
  \bibinfo{journal}{Phys. Rev. E} \textbf{\bibinfo{volume}{48}},
  \bibinfo{pages}{4351} (\bibinfo{year}{1993}).

\bibitem[{\citenamefont{Donev et~al.}(2005)\citenamefont{Donev, Torquato, and
  Stillinger}}]{Event_Driven_HE}
\bibinfo{author}{\bibfnamefont{A.}~\bibnamefont{Donev}},
  \bibinfo{author}{\bibfnamefont{S.}~\bibnamefont{Torquato}}, \bibnamefont{and}
  \bibinfo{author}{\bibfnamefont{F.~H.} \bibnamefont{Stillinger}},
  \bibinfo{journal}{J. Comp. Phys.} \textbf{\bibinfo{volume}{202}},
  \bibinfo{pages}{737} (\bibinfo{year}{2005}).

\bibitem[{\citenamefont{Eppenga and Frenkel}(1984)}]{VirtualScaling_MC}
\bibinfo{author}{\bibfnamefont{R.}~\bibnamefont{Eppenga}} \bibnamefont{and}
  \bibinfo{author}{\bibfnamefont{D.}~\bibnamefont{Frenkel}},
  \bibinfo{journal}{Mol. Phys.} \textbf{\bibinfo{volume}{52}},
  \bibinfo{pages}{1303} (\bibinfo{year}{1984}).

\bibitem[{\citenamefont{Gottschalk}(2000)}]{OBB_Thesis}
\bibinfo{author}{\bibfnamefont{S.}~\bibnamefont{Gottschalk}}, Ph.D. thesis,
  \bibinfo{school}{UNC Chapel Hill}, \bibinfo{address}{Department of Computer
  Science} (\bibinfo{year}{2000}).

\bibitem[{\citenamefont{Boubl\'{i}k}(1975)}]{SPT_convex_2D}
\bibinfo{author}{\bibfnamefont{T.}~\bibnamefont{Boubl\'{i}k}},
  \bibinfo{journal}{Mol. Phys.} \textbf{\bibinfo{volume}{29}},
  \bibinfo{pages}{421} (\bibinfo{year}{1975}).

\bibitem[{\citenamefont{Kenyon et~al.}(2000)\citenamefont{Kenyon, Propp, and
  Wilson}}]{RandomTiling_SpanningTree}
\bibinfo{author}{\bibfnamefont{R.~W.} \bibnamefont{Kenyon}},
  \bibinfo{author}{\bibfnamefont{J.~G.} \bibnamefont{Propp}}, \bibnamefont{and}
  \bibinfo{author}{\bibfnamefont{D.~B.} \bibnamefont{Wilson}},
  \bibinfo{journal}{Electronic J. of Combinatorics}
  \textbf{\bibinfo{volume}{7}}, \bibinfo{pages}{R25} (\bibinfo{year}{2000}).

\bibitem[{\citenamefont{Cohn et~al.}(2001)\citenamefont{Cohn, Kenyon, and
  Propp}}]{RandomTilings_Entropy}
\bibinfo{author}{\bibfnamefont{H.}~\bibnamefont{Cohn}},
  \bibinfo{author}{\bibfnamefont{R.}~\bibnamefont{Kenyon}}, \bibnamefont{and}
  \bibinfo{author}{\bibfnamefont{J.}~\bibnamefont{Propp}}, \bibinfo{journal}{J.
  Am. Math. Soc.} \textbf{\bibinfo{volume}{14}}, \bibinfo{pages}{297}
  (\bibinfo{year}{2001}).

\bibitem[{\citenamefont{Low}(2002)}]{BiaxialOrderMetrics}
\bibinfo{author}{\bibfnamefont{R.~J.} \bibnamefont{Low}},
  \bibinfo{journal}{Eur. J. Phys.} \textbf{\bibinfo{volume}{23}},
  \bibinfo{pages}{111} (\bibinfo{year}{2002}).

\bibitem[{\citenamefont{Frenkel and Eppenga}(1985)}]{LRO_HardRods_2D}
\bibinfo{author}{\bibfnamefont{D.}~\bibnamefont{Frenkel}} \bibnamefont{and}
  \bibinfo{author}{\bibfnamefont{R.}~\bibnamefont{Eppenga}},
  \bibinfo{journal}{Phys. Rev. A} \textbf{\bibinfo{volume}{31}},
  \bibinfo{pages}{1776} (\bibinfo{year}{1985}).

\bibitem[{\citenamefont{Bates and Frenkel}(2000)}]{SpherocyllindersPhases_2D}
\bibinfo{author}{\bibfnamefont{M.~A.} \bibnamefont{Bates}} \bibnamefont{and}
  \bibinfo{author}{\bibfnamefont{D.}~\bibnamefont{Frenkel}},
  \bibinfo{journal}{J. of Chem. Phys.} \textbf{\bibinfo{volume}{112}},
  \bibinfo{pages}{10034} (\bibinfo{year}{2000}).

\bibitem[{\citenamefont{Weber et~al.}(1995)\citenamefont{Weber, Marx, and
  Binder}}]{HardDiskMelting_Scaling}
\bibinfo{author}{\bibfnamefont{H.}~\bibnamefont{Weber}},
  \bibinfo{author}{\bibfnamefont{D.}~\bibnamefont{Marx}}, \bibnamefont{and}
  \bibinfo{author}{\bibfnamefont{K.}~\bibnamefont{Binder}},
  \bibinfo{journal}{Phys. Rev. B} \textbf{\bibinfo{volume}{51}},
  \bibinfo{pages}{14636} (\bibinfo{year}{1995}).

\bibitem[{\citenamefont{Richard et~al.}(1998)\citenamefont{Richard, Höffe,
  Hermisson, and Baake}}]{RandomTilings_Math}
\bibinfo{author}{\bibfnamefont{C.}~\bibnamefont{Richard}},
  \bibinfo{author}{\bibfnamefont{M.}~\bibnamefont{Höffe}},
  \bibinfo{author}{\bibfnamefont{J.}~\bibnamefont{Hermisson}},
  \bibnamefont{and} \bibinfo{author}{\bibfnamefont{M.}~\bibnamefont{Baake}},
  \bibinfo{journal}{J. Phys. A: Math. Gen.} \textbf{\bibinfo{volume}{31}},
  \bibinfo{pages}{6385} (\bibinfo{year}{1998}).

\bibitem[{\citenamefont{Wojciechowski}(1992)}]{DiskDimers_DC}
\bibinfo{author}{\bibfnamefont{K.~W.} \bibnamefont{Wojciechowski}},
  \bibinfo{journal}{Phys. Rev. B} \textbf{\bibinfo{volume}{46}},
  \bibinfo{pages}{26} (\bibinfo{year}{1992}).

\bibitem[{\citenamefont{Wojciechowski}(1987)}]{DiskDimers_FreeVolume}
\bibinfo{author}{\bibfnamefont{K.~W.} \bibnamefont{Wojciechowski}},
  \bibinfo{journal}{Phys. Lett. A} \textbf{\bibinfo{volume}{122}},
  \bibinfo{pages}{377} (\bibinfo{year}{1987}).

\bibitem[{\citenamefont{Kenyon}(1997)}]{DimerCovering_Statistics}
\bibinfo{author}{\bibfnamefont{R.}~\bibnamefont{Kenyon}},
  \bibinfo{journal}{Annales de Inst. H. Poincar\'e, Probabilit\'es et
  Statistiques} \textbf{\bibinfo{volume}{33}}, \bibinfo{pages}{591}
  (\bibinfo{year}{1997}).

\bibitem[{\citenamefont{Donev et~al.}(2004)\citenamefont{Donev, Torquato,
  Stillinger, and Connelly}}]{Jamming_LP_results}
\bibinfo{author}{\bibfnamefont{A.}~\bibnamefont{Donev}},
  \bibinfo{author}{\bibfnamefont{S.}~\bibnamefont{Torquato}},
  \bibinfo{author}{\bibfnamefont{F.~H.} \bibnamefont{Stillinger}},
  \bibnamefont{and} \bibinfo{author}{\bibfnamefont{R.}~\bibnamefont{Connelly}},
  \bibinfo{journal}{J. App. Phys.} \textbf{\bibinfo{volume}{95}},
  \bibinfo{pages}{989} (\bibinfo{year}{2004}).

\bibitem[{\citenamefont{Torquato et~al.}(2000)\citenamefont{Torquato, Truskett,
  and Debenedetti}}]{Torquato_MRJ}
\bibinfo{author}{\bibfnamefont{S.}~\bibnamefont{Torquato}},
  \bibinfo{author}{\bibfnamefont{T.~M.} \bibnamefont{Truskett}},
  \bibnamefont{and} \bibinfo{author}{\bibfnamefont{P.~G.}
  \bibnamefont{Debenedetti}}, \bibinfo{journal}{Phys. Rev. Lett.}
  \textbf{\bibinfo{volume}{84}}, \bibinfo{pages}{2064} (\bibinfo{year}{2000}).

\bibitem[{\citenamefont{Cuesta and Frenkel}(1990)}]{EllipsesPhases_MC}
\bibinfo{author}{\bibfnamefont{J.~A.} \bibnamefont{Cuesta}} \bibnamefont{and}
  \bibinfo{author}{\bibfnamefont{D.}~\bibnamefont{Frenkel}},
  \bibinfo{journal}{Phys. Rev. A} \textbf{\bibinfo{volume}{42}},
  \bibinfo{pages}{2126} (\bibinfo{year}{1990}).

\end{thebibliography}

\end{document}